\def\year{2022}\relax
\documentclass[letterpaper]{article} 
\usepackage{aaai22}  
\usepackage{times}  
\usepackage{helvet} 
\usepackage{courier}  
\usepackage[hyphens]{url}  
\usepackage{graphicx} 
\usepackage{graphicx}  
\usepackage{natbib}  
\usepackage{caption} 
\usepackage{xr} 
\usepackage{amsmath}
\usepackage{amssymb}
\usepackage{amsthm}
\usepackage{xcolor}
\usepackage[hidelinks]{hyperref}
\usepackage{graphicx}
\usepackage{booktabs}
\usepackage{algorithm}
\usepackage{algorithmic}
\usepackage{multirow}
\usepackage{scrextend}
\usepackage{lscape} 
\newcommand{\ohist}{\Omega^o_t}

\newcommand{\lastobid}{\omega^o_{t}}
\newcommand{\accbid}{\omega^{acc}}
\newcommand{\uthresh}{\bar{u}_t}
\newcommand{\param}[1]{{\color{blue}#1}}
\newcommand{\estUsermodel}{\widehat{U}_u}
\newcommand{\estOppmodel}{\widehat{U}_o}
\newcommand{\realUsermodel}{U_u}

\renewcommand{\subsubsection}[1]{\vspace*{5pt}\noindent\textbf{#1}\hspace*{2pt}}
\renewcommand{\paragraph}[1]{\vspace*{5pt}\noindent\textbf{#1}\hspace*{2pt}}

\title{Learnable Strategies for Bilateral Agent Negotiation over Multiple Issues}
\author{Pallavi Bagga$^1$ , Nicola Paoletti$^2$ , Kostas Stathis$^3$  \\ Royal Holloway, University of London, UK \\ \{pallavi.bagga.2017$^{1}$, nicola.paoletti$^{2}$, kostas.stathis$^{3}$\}@rhul.ac.uk}

\begin{document}

\maketitle

\section*{Abstract}
We present a novel bilateral agent negotiation model supporting multiple issues under user preference uncertainty. This model relies upon interpretable {\em strategy templates} representing agent negotiation tactics with learnable parameters. It also uses deep reinforcement learning to evaluate threshold utilities for those tactics that require them. To handle user preference uncertainty, our approach uses stochastic search to derive the user model that best approximates a given partial preference profile. Multi-objective optimization and multi-criteria decision-making methods are applied at negotiation time to generate near Pareto-optimal bids. We empirically show that our agent model outperforms winning agents of previous automated negotiation competitions in terms of the individual as well as social-welfare utilities. Also, the adaptive nature of our learning-based agent model provides a significant advantage when our agent confronts unknown opponents in unseen negotiation scenarios.

\section{Introduction}
An important problem in multi-issue bilateral negotiation is how to model a self-interested agent learn to adapt its strategy while negotiating with other agents. A model of this kind mostly considers the application preferences of the user the agent represents. Think, for instance, of bilateral negotiation in e-commerce, 
where a buyer agent settles 
the product price according to user preferences such as product colour, payment method and delivery time~\cite{fatima2006multi}. In practice, users would express their preferences by ranking only a few representative examples instead of specifying a utility function~\cite{tsimpoukis2018automated}. Hence, agents are uncertain about the complete user preferences and lack knowledge about the preferences and characteristics of their opponents
~\cite{baarslag2016learning}. 
In such uncertain settings, predefined one-size-fits-all heuristics are unsuitable for representing a strategy, which must learn from opponent actions in different domains. 

To this aim, we introduce \textit{ANESIA} (\textit{A}daptive \textit{NE}gotiation model for a \textit{S}elf-\textit{I}nterested \textit{A}utonomous agent), an agent model that provides a learnable, adaptive and interpretable strategy under user and opponent preferences' uncertainty. \textit{ANESIA} builds on so-called \textit{strategy templates}, i.e., parametric strategies that incorporate multiple negotiation tactics for the agent to choose from. A strategy template is described by a set of condition-action rules to be applied at different stages during the negotiation. Templates require no assumptions from the agent developer as to which tactic to choose during negotiation: from a template, we automatically learn 
 the best combination of tactics 
to use at any time during the negotiation. 
Being logical combinations of individual tactics, the resulting strategies are interpretable and, thus, can be explained to the user. Although the template parameters are learned before a negotiation begins, \textit{ANESIA} allows for online learning and adaptation as well. 

\textit{ANESIA} is 
an actor-critic architecture combining Deep Reinforcement Learning (DRL)~\cite{lillicrap2017continuous} with 
meta-heuristic optimisation.
DRL allows \textit{ANESIA} to learn a near-optimal threshold utility dynamically, adapting the strategy in different domains and against unknown opponents. \textit{ANESIA} neither accepts nor proposes any bid below this utility.
To deal with user preference uncertainty, we use a single-objective 
meta-heuristic
~\cite{yang2020nature, yang2009firefly} that handles 
large search spaces optimally,
and converges towards a good solution in a limited number of iterations, using a limited population size~\cite{talbi2009metaheuristics}.
This estimated user model  
ensures that our agent receives a satisfactory utility from opponent bids. 
Also, \textit{ANESIA} generates near-Pareto-optimal bids leading to win-win situations by combining multi-objective 
optimization (MOO)
~\cite{deb2002fast} and multi-criteria decision-making ~\cite{hwang1981methods} on top of   estimated user and opponent~models. 

To evaluate the effectiveness of our approach against the state-of-the-art, we conduct simulation experiments based on the 
ANAC tournaments~\cite{jonker2017automated}.
We have chosen domains 
available in the GENIUS tool~\cite{lin2014genius},
with different domain sizes and competitiveness levels~\cite{williams2014overview}. We play against winning agents with learning capabilities (from ANAC'17 and '18) and agents that deal with user preference uncertainty (from ANAC'19). These agents span a wide range of strategies and techniques\footnote{E.g., Genetic algorithm-\textit{SAGA}, Bayesian approach-\textit{FSEGA}, Gaussian process-\textit{AgentGP}, Tabu Search-\textit{KakeSoba}, Logistic Regression-\textit{AgentHerb}, and Statistical frequency model-\textit{AgentGG}.}.
Empirically, \textit{ANESIA} outperforms existing strategies in terms of individual and social welfare utilities.

\paragraph{Related work} Existing approaches for a bilateral multi-issue strategy with reinforcement learning have focused on Tabular Q-learning to learn what to bid~\cite{bakker2019rlboa} or DQN to learn when to accept a bid~\cite{razeghi2020deep}.
Unlike our work, these approaches are neither optimal for continuous action spaces nor can handle user preference uncertainty. 
The negotiation model of~\cite{bagga2020deep} uses DRL to learn a complete strategy, but it is for single issue only. All the above approaches use DRL to learn acceptance or bidding strategies. Instead we learn a threshold utility as part of (possibly more sophisticated) tactics for acceptance and bidding.

Past work uses meta-heuristics to explore the outcome space and find desired bids or for bid generation~\cite{silva2018optimizing,de2016gangster,el2020chaotic,sato2016whaleagent,kadono2016agent,klein2003negotiating, sato2016whaleagent}. 
We, too, use population-based meta-heuristic, but for estimating the user model that best agrees with the partial user preferences.
In particular, we employ the Firefly Algorithm (FA)~\cite{yang2009firefly} that has shown its effectiveness in continuous optimization, but has not been tested until now in the context of automated negotiation.
Moreover, while the Genetic Algorithm NSGA-II~\cite{deb2002fast} for 
MOO
has been used previously to find multiple Pareto-optimal solutions during negotiation~\cite{hashmi2013automated}, we are the first to combine NSGA-II with TOPSIS~\cite{hwang1981methods} to choose one best among a set of ranked near Pareto-optimal outcomes. 

\vspace{-2mm}
\section{Negotiation Environment}\label{NegEnv}
We assume \textit{bilateral negotiations} where two agents interact with each other in a domain $D$, over $n$ different \textit{independent} issues, $D = (I_1, I_2,  \dots I_n)$, with each issue taking a finite set of $k$ possible discrete or continuous values $I_i = (v^i_1, \ldots v^i_{k})$.  
In our experiments, 
we consider issues with discrete values. An agent's bid $\omega$ is a mapping from each issue to a chosen value (denoted by $c_i$ for the $i$-th issue), i.e., $\omega = (v^1_{c_1}, \ldots v^n_{c_n})$.  
The set of all possible bids or outcomes is called outcome space
$\Omega$ s.t. $\omega \in \Omega$. The outcome space is common knowledge to the negotiating parties and stays fixed during a single negotiation session. 

\paragraph{Negotiation protocol} Before the agents can begin the negotiation and exchange bids, they must agree on a negotiation protocol $P$, which determines the valid moves agents can take at any state of the negotiation ~\cite{fatima2005comparative}. Here, we consider the \textit{alternating offers protocol}
~\cite{rubinstein1982perfect} due to its simplicity and wide use. The set of protocol's possible $Actions$ are: $\{\mathit{offer}(\omega), \mathit{accept}, \mathit{reject}\}$. 
One of the agents (say $A_u$) starts a negotiation by making an offer $X$ to the other agent (say $A_o$). Agent $A_o$ can either accept or reject the offer. If it accepts, the negotiation ends with an agreement, otherwise $A_o$ makes a counter-offer to $A_u$. This process of making offers continues until one of the agents either accepts an offer (i.e., success) or the deadline is reached (i.e., failure). 
Moreover, we assume that the negotiations are sensitive to \textit{time}, i.e. time impacts the utilities of the negotiating parties. In other words, the value of an agreement decreases over time.

\paragraph{Utility} 
Each agent has a preference profile, reflecting the issues important to the agent, which values per issue are preferred over
other values, and on the whole provides a (partial) ranking over all possible deals~\cite{marsa2014problems}. In contrast to $\Omega$, the agent's preference profile is private information and is given in terms of a utility function $U$. $U$ is defined as a weighted sum of evaluation functions, $e_i(v^i_{c_i})$ as shown in \eqref{utilitySpace}. 
Each issue is evaluated separately and contributes linearly to
$U$. 
Such a $U$ is a very common utility model and is also called a \textit{Linear Additive Utility space}. Here, $w_i$ are the normalized weights indicating
each issue's importance
to the user and $e_i(v^i_{c_i})$ is an evaluation function that maps the $v^i_{c_i}$ value of the $i^{th}$ issue to a utility. 
\vspace*{-\parskip}
\begin{equation}
\label{utilitySpace}
\resizebox{.99\linewidth}{!}{$
  U(\omega) = U (v^1_{c_1}, \ldots v^n_{c_n}) = \sum^n_{i=1} w_i \cdot e_i(v^i_{c_i}), \text{ where } \sum^n_{i=1} w_i = 1 
$}
\end{equation}
Note that 
$U$
does not take dependencies between issues into account. Whenever the negotiation terminates without any agreement, each negotiating party gets its corresponding utility based on the private reservation\footnote{The reservation value is the minimum acceptable utility for an agent. It may vary for different parties and different domains. In our settings, it is the same for both parties.} value ($u_{res}$). In case the negotiation terminates with an agreement, each agent receives the discounted utility of the agreed bid, i.e., $U^d(\omega) = U (\omega)d_D^t$. Here, $d_D$ is a discount factor in the interval $[0,1]$ and $t \in [0,1]$ is current normalized time.

\paragraph{User and opponent utility models} 
In our settings, the negotiation environment is one with \textit{incomplete information}, because the user utility model $U_u$ is unknown. Only partial preferences are given, i.e., a
partial order $\preceq$
over $B$ bids w.r.t.\ $U_u$ 
s.t.\ $\omega_1 \preceq \omega_2 \rightarrow U_u(\omega_1) \leq U_u(\omega_2)$. Hence, during the negotiation, one of the objectives of our agent is to derive an estimate $\estUsermodel$ of the real utility function $U_u$ from the given partial preferences\footnote{Humans do not necessarily use an explicit utility function. 
Also, preference elicitation can be tedious for users since they have to interact with the system repeatedly~\cite{baarslag2017value}. As a result, agents should accurately represent users under minimal preference information~\cite{tsimpoukis2018automated}.}.
This leads to a single-objective constrained optimization problem,
\vspace{-2mm}
\begin{equation}
    \label{singleobjective}
    \begin{aligned}
    \underset{\stackrel{\widehat{w_1},\ldots,\widehat{w_n},}{ \widehat{e_1}(v^1_{c_1}),\ldots, \widehat{e_n}(v^n_{c_n})}}{\max} \quad & 
    \rho\left(\sum^n_{i=1} \widehat{w_i} \cdot \widehat{e_i}(v^i_{c_i}), B_{\preceq}\right) \\
    \textrm{s. t.} \quad \quad & \sum_{i=1}^{n}{\widehat{w_i}} = 1 \\
    & \widehat{w_i} > 0 \text{ and } 0 \leq \widehat{e_i}(v^i_{c_i}), \forall i \in n
    \end{aligned}
\end{equation}
where $B_{\preceq}$ is the incomplete sequence of known bid preferences (ordered by $\preceq$), and  $\rho$ is a measure of ranking similarity (e.g., Spearman correlation) between the estimated ranking of $\widehat{U}_u$ and the true, but partial, bid ranking $B_{\preceq}$.
We also assume that our agent is unaware of the utility structure of its opponent agent $U_o$. Hence, to increase the agreement rate over multiple issues, another objective of our agent is to generate the (near) Pareto-optimal solutions during the negotiation which can be defined as a MOO problem as follows:
\begin{equation}
\label{multiobjective}
\begin{aligned}
 \underset{\omega \in \Omega}{\max} \quad & (\estUsermodel (\omega), \estOppmodel (\omega)) 
\end{aligned}
\end{equation}
\noindent In \eqref{multiobjective}, we have two objectives: $\widehat{U}_u$, the user's estimated utility, and $\estOppmodel$, the opponent's estimated utility. 
A bid $\omega^* \in \Omega$ is Pareto optimal if no other bid exists $\omega \in \Omega$ that Pareto-dominates $\omega^*$. In our case, a bid $\omega_1$ Pareto-dominates $\omega_2$ iff:
\begin{equation}
\label{paretobid}
\begin{aligned}
    \left(\estUsermodel(\omega_1) \geq \estUsermodel(\omega_2) \land \estOppmodel(\omega_1) \geq \estOppmodel(\omega_2)\right) \land \\
    \left(\estUsermodel(\omega_1) > \estUsermodel(\omega_2) \lor \estOppmodel(\omega_1) > \estOppmodel(\omega_2)\right)
\end{aligned}
\end{equation}

\section{The \textit{ANESIA} Model}\label{proposedmodel}
Our agent $A_u$ is situated in an environment $E$ (containing the opponent agent $A_o$) where at any time $t$, $A_u$ senses the current state $S_t$ of $E$ and represents it as a set of internal attributes.
These include information derived from the sequence of previous bids offered by $A_o$  (e.g., utility of the best opponent bid so far $O_{best}$, average utility of all the opponent bids $O_{avg}$ and their variability $O_{sd}$) and information stored in our agent's knowledge base (e.g., number of bids $B$ in the given partial order, $d_D$, $u_{res}$, $\Omega$, and $n$), and the current negotiation time $t$.
This internal state representation, denoted with $s_t$, is used by the agent (in acceptance and bidding strategies) to decide what action $a_t$ to execute. 
Action execution then changes the state of the environment to $S_{t+1}$.

Learning in \textit{ANESIA}\footnote{See Appendix for the interaction between components of \textit{ANESIA} architecture.} mainly consists of three components: \textit{Decide}, \textit{Negotiation Experience}, and \textit{Evaluate}. 
\textit{Decide} refers to the negotiation strategy for choosing a near-optimal action $a_t$ among a set of $Actions$ at a particular state $s_t$ based on a protocol $P$. Action
$a_t$
is derived via two functions, {$f_a$} and {$f_b$}, for the acceptance and bidding strategies, respectively. Function $f_a$ takes as inputs $s_t$, a \textit{dynamic threshold utility} $\bar{u}_t$ (defined later in the Methods section), the sequence of past opponent bids $\ohist$,
and outputs a discrete action $a_t$ among \textit{accept} or \textit{reject}. When $f_a$ returns \textit{reject}, $f_b$ computes what to bid next, with input $s_t$ and $\bar{u}_t$, see~(\ref{eq:acceptance}--\ref{eq:bidding}). 
This separation of acceptance and bidding strategies 
is not rare, see for instance~\cite{baarslag2014decoupling}.
\begin{align}
    f_a(s_t, \bar{u}_t, \ohist) = & \ a_t, a_t \in \{\mathit{accept, reject}\}\label{eq:acceptance}\\
        f_b(s_t,\bar{u}_t, \ohist) = & \ a_t, a_t \in \{\mathit{offer}(\omega), \omega \in \Omega \}
        \label{eq:bidding}
\end{align}
Since we assume incomplete user and opponent preference information, \textit{Decide} uses the estimated models $\estUsermodel$ and $\estOppmodel$. In particular, 
$\estUsermodel$ is estimated once before the negotiation starts by solving \eqref{singleobjective} and using the given partial preference profile $\preceq$.
This encourages agent autonomy and avoids continuous user preference elicitation. Similarly, $\estOppmodel$ is estimated at time $t$ using information from $\ohist$, see Methods section for more details. 

\textit{Negotiation Experience} stores historical information about $N$ previous
interactions of an agent with other agents. 
Experience elements are of the form $\langle s_t, a_t, r_t, s_{t+1} \rangle$, where $s_t$ is the internal state representation of 
the negotiation environment 
$E$, 
$a_t$ is the performed action, 
$r_t$ is a scalar \textit{reward} received from the environment and $s_{t+1}$ is the new agent state after executing $a_t$. 

\textit{Evaluate} refers to a critic helping {\em ANESIA} learn the dynamic threshold utility $\bar{u}_t$, which evolves as new experience is collected. 
More specifically, it is a function of random $K$ ($K<N$) experiences fetched from the agent's memory. Learning $\bar{u}_t$ is \textit{retrospective} since it depends on the {reward} $r_t$ obtained from $E$ by performing $a_t$ at $s_t$. The reward value depends on the (estimated) discounted utility of the last bid received from the opponent, $\lastobid$, or of the bid accepted by either parties $\accbid$ and defined as follows:

\begin{equation}\label{reward}
    r_t = 
    \begin{cases}
    \estUsermodel(\accbid,t), & \text{on agreement} \\
    \estUsermodel(\lastobid,t), & \text{on received offer} \\
    -1, & \text{otherwise}.
    \end{cases}
\end{equation}
$\estUsermodel(\omega,t)$ is the discounted reward of $\omega$ defined as:
\begin{equation}\label{utility}
   \estUsermodel(\omega,t) =\widehat{U}(\omega)\cdot{d^t}, d \in [0,1]
\end{equation}
where $d$ is a temporal discount factor to encourage the agent to negotiate without delay. 
We should not confuse $d$, which is typically unknown to the agent, with the discount factor used to compute the utility of an agreed bid ($d_D$).

\paragraph{Strategy templates:} One common way to define the acceptance ($f_a$) and bidding ($f_b$) strategies is via a combination of hand-crafted tactics that, by empirical evidence or domain knowledge, are known to work effectively. However, a fixed set of tactics might not well adapt to multiple different negotiation domains. {\em ANESIA} does not assume pre-defined strategies for $f_a$ and $f_b$, and learns these strategies \textit{offline}. We run multiple negotiations between our agent and a pool of opponents. We select the combination of tactics that maximizes the \textit{true} user utility over these negotiations. So, in this stage only, we assume that the true user model is known.

To enable strategy learning, we introduce
\textit{strategy templates}, i.e., parametric strategies incorporating a series of tactics, where each tactic is executed for a specific negotiation phase. The parameters describing the start and duration of each phase, as well as the particular tactic choice for that phase are all \textit{learnable} (blue-colored in~\eqref{eq:acc_templ}, \eqref{eq:bid_templ}). Moreover, tactics can expose, in turn, learnable parameters themselves. 

We assume a 
collection of acceptance and bidding tactics, $\mathcal{T}_a$ and $\mathcal{T}_b$. Each $\mathtt{t}_a \in \mathcal{T}_a$ maps the agent state, threshold utility, opponent bid history, and a (possibly empty) vector of learnable parameters $\mathbf{p}$ into a utility value: if the agent is using tactic $\mathtt{t}_a$ and $\mathtt{t}_a(s_t, \bar{u}_t, \ohist, \mathbf{p}) = u$, then it will not accept any offer with utility below $u$, see \eqref{eq:acc_templ} below. 
Each $\mathtt{t}_b \in \mathcal{T}_b$ is of the form $\mathtt{t}_b(s_t, \bar{u}_t, \ohist, \mathbf{p}) = \omega$ where $\omega\in \Omega$ is the bid returned by the tactic. 
An \textit{acceptance strategy template} is a parametric function given by
\vspace*{-\parskip}
\begin{equation}\label{eq:acc_templ}
\resizebox{.99\linewidth}{!}{$
    \bigwedge_{i=1}^{n_a} t \in [t_i, t_{i+1}) \rightarrow \left( \bigwedge_{j=1}^{n_i}
  { \color{blue}{c_{i,j}}} \rightarrow \widehat{U}(\lastobid) \geq \mathtt{t}_{i,j}(s_t, \bar{u}_t, \ohist, \param{\mathbf{p}_{i,j}})\right)$}
\end{equation}
where $n_a$ is the number of phases; $t_1=0$, $t_{n_a+1}=1$, and $t_{i+1}=t_i+ \param{\delta_i}$, where the $\param{\delta_i}$ parameter determines the duration of the $i$-th phase; 
for each phase $i$, the strategy template includes $n_i$ tactics to choose from: $\param{c_{i,j}}$ is a Boolean choice parameter determining whether tactic $\mathtt{t}_{i,j}\in \mathcal{T}_a$ should be used during the $i$-th phase. 
We note that~\eqref{eq:acc_templ} is a predicate returning whether or not the opponent bid $\lastobid$ is accepted. Similarly, 
a \textit{bidding strategy template} is defined by 
\vspace*{-\parskip}
\begin{equation}\label{eq:bid_templ}
\resizebox{.99\linewidth}{!}{$
    \bigcup_{i=1}^{n_b} 
    \begin{cases}
    \mathtt{t}_{i,1}(s_t, \bar{u}_t, \ohist, \param{\mathbf{p}_{i,1}}) & \text{ if } t \in [t_i, t_{i+1})  \text{ and } \param{c_{i,1}}\\
    \cdots & \cdots\\
    \mathtt{t}_{i,n_{i}}(s_t, \bar{u}_t, \ohist,  \param{\mathbf{p}_{i,n_i}}) & \text{ if } t \in [t_i, t_{i+1})  \text{ and } \param{c_{i,n}}
    \end{cases}$}
\end{equation}
where $n_b$ is the number of phases, $n_i$ is the number of options for the $i$-th phase, and $\mathtt{t}_{i,j}\in \mathcal{T}_b$. $t_i$ and $\param{c_{i,j}}$ are defined as in the acceptance template. The particular libraries of tactics used in this work are discussed in the next Section. 
We stress that both \eqref{eq:acc_templ} and~\eqref{eq:bid_templ} describe time-dependent strategies where a given choice of tactics is applied at different phases (denoted by the condition $t \in [t_i, t_{i+1})$).
\section{Methods}\label{methods}
\vspace{-2mm}
\paragraph{User modelling:}
Before the negotiation begins, we estimate the user model $\estUsermodel$ by finding the weights $w_i$ and utility values $e_i(v_{c_i}^i)$ for each issue $i$, see~\eqref{utilitySpace}, so that the resulting bid ordering best fits the given partial order $\preceq$ of bids. 
To solve this optimization problem~\eqref{singleobjective}, 
we use FA~\cite{yang2009firefly}, a meta-heuristic inspired by the swarming and flashing behaviour of 
fireflies, because, in our preliminary analyses, it outperformed other traditional nature-inspired meta-heuristics such as GA and PSO~\cite{ritu2021}. 
We compute the fitness of a candidate solution (i.e., the user model $\estUsermodel'$) as the Spearman's rank correlation coefficient $\rho$ between the estimated ranking of $\estUsermodel'$ and the true, but partial, bid ranking $\preceq$. The coefficient $\rho \in [-1,1]$ is indeed a similarity measure between two rankings, assigning a value of $1$ for identical and $-1$ for opposed rankings.  

\paragraph{Opponent modelling:}
To derive an estimate of the opponent model $\estOppmodel$ during negotiation, 
we use the distribution-based frequency model proposed in~\cite{tunali2017rethinking}.
In this model, the empirical frequency of the issue values in $\ohist$ provides an educated guess on the opponent's most preferred issue values. 
The issue weights are estimated by analysing the disjoint windows of $\ohist$, giving an idea of the shift of opponent's preferences from its previous negotiation strategy over time.

\paragraph{Utility threshold learning:}
We use an actor-critic architecture with model-free deep reinforcement learning (i.e., Deep Deterministic Policy Gradient (DDPG)~\cite{lillicrap2017continuous}) to predict the target threshold utility $\uthresh$. We consider a model-free RL approach because our problem is how to make an agent decide what target threshold utility to set next in a negotiation dialogue rather than predicting the new state of the environment, which implies model-based RL. Thus, $\uthresh$ is expressed as a deep neural network function, which takes the agent state $s_t$ as an input (see previous section for the list of attributes). Prior to RL, our agent's strategy is pre-trained with supervision from synthetic negotiation data. 
To collect supervision data, we use the \textit{GENIUS} simulation environment~\cite{lin2014genius}, which supports multi-issue bilateral negotiation for different domains and user profiles. In particular, data was generated by running the winner of the ANAC'19
(AgentGG) against other strategies\footnote{\textit{Gravity}, \textit{HardDealer}, \textit{Kagent}, \textit{Kakesoba}, \textit{SAGA}, and \textit{SACRA}.} in three different domains\footnote{\label{domains}Laptop, Holiday and Party.} and 
assuming no user preference uncertainties~\cite{aydougan2020challenges}. This initial supervised learning (SL) stage helps our agent decrease the exploration time required for DRL during the negotiation, an idea primarily influenced by the work of~\cite{bagga2020deep}. 

\paragraph{Strategy learning:} 
The parameters of the acceptance and bidding strategy templates  (\ref{eq:acc_templ}--\ref{eq:bid_templ}) are learned by running the FA meta-heuristic.
We define the fitness of a particular choice of template parameters as the average \textit{true} user utility over multiple negotiations rounds under the concrete strategy implied by those parameters.
Negotiation data is obtained by running our agent on the GENIUS platform against three (readily available) opponents (\textit{AgentGG}, \textit{KakeSoba} and
\textit{SAGA}) in three different negotiation domains\footref{domains}.

We now describe the libraries of tactics used in our templates. As for the acceptance tactics, we consider: 
\newcommand{\myitem}{\noindent $\bullet$~}
\begin{itemize}
\item $\estUsermodel(\omega_t)$, the estimated utility of the bid that our agent would propose at the time $t$ ($\omega_t = f_b(s_t,\uthresh,\ohist)$). 
\item 
$Q_{\estUsermodel(\ohist)}(\param{a}\cdot t + \param{b})$, where $\estUsermodel(\ohist)$ is the distribution of (estimated) utility values of the bids in $\ohist$, $Q_{\estUsermodel(B_o(t))}(p)$ is the quantile function of such distribution, and $\param{a}$ and $\param{b}$ are learnable parameters. In other words, we consider the $p$-th best utility received from the agent, where $p$ is a learnable (linear) function of the negotiation time $t$. In this way, this tactic automatically and dynamically decides how much the agent should concede at time $t$.

\item $\uthresh$, the dynamic DRL-based utility threshold.

\item $\param{\bar{u}}$, a fixed, but learnable, utility threshold.
\end{itemize}
The bidding tactics in our library are:
\begin{itemize}
\item $b_{\mathit{Boulware}}$, a bid generated by a time-dependent Boulware strategy~\cite{fatima2001optimal}.
\item $PS(\param{a}\cdot t + \param{b})$ extracts a bid from the set of Pareto-optimal bids $PS$ (see \eqref{paretobid}), derived using the \textit{NSGA-II algorithm}\footnote{Meta-heuristics (instead of brute-force) for Pareto-optimal solutions have the potential to deal efficiently with continuous issues.}~\cite{deb2002fast} under 
$\estUsermodel$ and $\estOppmodel$.
In particular, it selects the bid that assigns a weight of $\param{a}\cdot t + \param{b}$ to our agent utility (and $1-(\param{a}\cdot t + \param{b})$ to the opponent's), where $\param{a}$ and $\param{b}$ are learnable parameters telling how this weight scales with the negotiation time $t$. The \textit{TOPSIS algorithm}~\cite{hwang1981methods}  is used to derive such a bid, given the weighting $\param{a}\cdot t + \param{b}$ as input. 
\item $b_{opp}(\lastobid)$, a tactic to generate a bid by manipulating the last bid received from the opponent $\lastobid$. This is modified in a greedy fashion by randomly changing the value of the least relevant issue (w.r.t.\ $\widehat{U}$) of $\lastobid$. 
\item $\omega \sim \mathcal{U}(\Omega_{\geq \uthresh})$, a random bid above our DRL-based utility threshold $\uthresh$\footnote{$\mathcal{U}(S)$ is the uniform distribution over $S$, and $\Omega_{\geq \uthresh}$ is the subset of $\Omega$ whose bids have estimated utility above $\uthresh$ w.r.t.\ $\widehat{U}$.}.
\end{itemize}

\noindent Below, we give an example of a concrete acceptance strategy learned in our experiments: it employs 
time-dependent quantile tactic during the middle of the negotiation, and the DRL threshold utility during the initial and final stages.
\begin{align*}
t \in [0.0, 0.4) \rightarrow & \
\widehat{U}(\lastobid) \geq \uthresh \wedge \bar{u} \\
t \in [0.4, 0.7) \rightarrow & \
\widehat{U}(\lastobid) \geq \widehat{U}(\omega_t) \wedge 
Q_{\widehat{U}(\ohist)}(-0.67 \cdot t + 1.27)\\
t \in [0.7, 0.95) \rightarrow & \
\widehat{U}(\lastobid) \geq \widehat{U}(\omega_t) \wedge 
Q_{\widehat{U}(\ohist)}(-0.21 \cdot t + 0.9) \\ 
t \in [0.95, 1.0] \rightarrow & \
\widehat{U}(\lastobid) \geq \uthresh
\end{align*} 
Below is an example of a learned concrete bidding strategy: it behaves in a Boulware-like manner in the initial stage, after which it proposes near Pareto-optimal bids (between time 0.4 and 0.9) and  opponent-oriented bid in the final stage. 
\begin{align*}
t \in [0.0, 0.4) \rightarrow & \
\omega = b_{\mathit{Boulware}} \\
t \in [0.4, 0.9) \rightarrow & \
\omega = PS(-0.75 \cdot t + 0.6) \\
t \in [0.9, 1.0] \rightarrow & \
\omega = b_{opp}(\lastobid)  
\end{align*}
We stress that our approach allows to automatically devise such combinations of tactics so as to achieve optimal user utility, which would be infeasible manually.

\begin{table}[h]
    \centering
      \resizebox{.48\textwidth}{!}{
    \begin{tabular}{p{0.13\textwidth}| p{0.07\textwidth}p{0.08\textwidth}| p{0.08\textwidth}p{0.08\textwidth}}
    \hline
    Domain (n) & \multicolumn{2}{|c|}{Ordinal Accuracy $(\uparrow)$} & \multicolumn{2}{c}{Cardinal Inaccuracy $(\downarrow)$} \\
    \hline
    $|B|$ &  $5\%$ of $\Omega$ & $10\%$ of $\Omega$  & $5\%$ of $\Omega$ & $10\%$ of $\Omega$ \\
    \hline
    AirportSite (3) & (0.75,0.75) & (0.85,0.87)  &  (0.47,0.54)  &  (0.78,0.76)\\
    Camera (6) &  (0.77,0.63) & (0.83,0.75) & (0.32,0.32) & (0.69,0.41) \\
    Energy (6) &  (0.74,0.78)  &  (0.83,0.84)  & (0.56,0.61)  &  (0.57,0.69) \\
    Fitness (5) & (0.67,0.67)  &  (0.70,0.75)  &  (0.55,0.47)  &  (0.46,0.59) \\
    Flight (3) & (0.75,0.85)  &  (0.82,0.90)  &  (0.65,0.75)  &  (0.58,0.79)\\
    Grocery (5) &  (0.67,0.67)  &  (0.75,0.72)  &  (0.35,0.42)  &  (0.39,0.56)  \\
    Itex-Cypress (4) &  (0.70,0.74)  & (0.78,0.80)  & (0.56,0.48)  &  (0.74,0.56) \\
    Outfit (4) &  (0.70,0.75)  & (0.80,0.84)  &  (0.71,0.88)  &  (0.89,0.79)\\
    \hline
    \end{tabular}}
    \caption{Evaluation of User Modelling using FA for two profiles (separated by comma) in each domain}
    \label{table:usermodel}
\end{table}
\vspace{-5mm}
\begin{table}[h]
    \centering
       \resizebox{.50\textwidth}{!}{
    \begin{tabular}{p{0.2\textwidth}p{0.07\textwidth}|
    p{0.2\textwidth}p{0.07\textwidth}}
    \hline
       Domain & IGD $(\downarrow)$ &  Domain & IGD $(\downarrow)$  \\
      \hline
    Airport Site ($|\Omega| = 420$) &  0.000   & Flight ($|\Omega| = 48$) & 0.006 \\
    Camera ($|\Omega| = 3600$) &  0.000  & Grocery ($|\Omega| = 1600$) &   0.000  \\
    Energy ($|\Omega| = 15625$) &  0.011 &  Itex-Cypress ($|\Omega| = 180$) & 0.000  \\
    Fitness ($|\Omega| = 3520$) &  0.012   &  Outfit ($|\Omega| = 128$) &  0.000\\
    \hline
    \end{tabular}
     }
    \caption{Evaluation of Pareto Frontier using Inverted Generational Distance estimated using NSGA-II}
    \label{table:paretoIGD}
    \vspace{-4mm}
\end{table}
\section{Experimental Results}\label{results}
All the experiments have been performed using the  GENIUS negotiation platform~\cite{lin2014genius}.
Our experiments are designed to prove the following hypotheses:

\noindent\textbf{Hypothesis A:} Our approach can 
well approximate
user models under user preference uncertainty.

\noindent\textbf{Hypothesis B:} The set of NSGA-II estimated Pareto-optimal bids are close to the true Pareto-optimal front. 

\noindent\textbf{Hypothesis C:} \textit{ANESIA} outperforms the ``teacher'' strategies (AgentGG, KakeSoba and SAGA) in known negotiation settings in terms of individual and social efficiency.

\noindent\textbf{Hypothesis D:} \textit{ANESIA} outperforms not-seen-before negotiation strategies and adapts to different negotiation settings in terms of individual and social efficiency.
\begin{table*}
    \centering
    \resizebox{.80\textwidth}{!}{
    \begin{tabular}{p{0.10\textwidth} p{0.16\textwidth}p{0.1\textwidth}
    p{0.1\textwidth}p{0.1\textwidth}
    p{0.1\textwidth}p{0.06\textwidth}}
    \hline
      Agent &
      $R_{\it avg} (\downarrow)$  &  
      $P_{\it avg} (\downarrow)$  &  
      $U_{\it soc} (\uparrow)$  &  
      $U_{\it ind}^{total} (\uparrow)$  &  
      $U_{\it ind}^s (\uparrow)$  &
      $S_{\%} (\uparrow)$ \\  
      \hline
      ANESIA & 301.64 $\pm$ 450.60 & \textbf{0.12 $\pm$ 0.36} & \textbf{1.55 $\pm$ 0.73} & 0.39 $\pm$ 0.41 & \textbf{0.92 $\pm$ 0.11} & 0.46 \\
      AgentGG & 1327.51 $\pm$ 2246.83 & 0.30 $\pm$ 0.37 & 1.10 $\pm$ 0.67 & 0.65 $\pm$ 0.39 & 0.87 $\pm$ 0.13 & 0.75 \\
      KakeSoba & 1154.83 $\pm$ 2108.11 & 0.22 $\pm$ 0.34 & 1.26 $\pm$ 0.62 & \textbf{0.71 $\pm$ 0.35} & 0.88 $\pm$ 0.12 & 0.81 \\
      SAGA & \textbf{287.34 $\pm$ 1058.52} & 0.18 $\pm$ 0.31 & 1.36 $\pm$ 0.56 & 0.58 $\pm$ 0.26 & 0.67 $\pm$ 0.14 & \textbf{0.87 }\\
         \hline
    \end{tabular}
    }
    \caption{Performance Comparison of \textit{ANESIA} with ``teacher'' strategies. Results are averaged for all the domains and profiles. See Appendix for separate results for each domains.}
    \label{table:teacher}
    \vspace{-4mm}
\end{table*}

\noindent\textbf{Performance metrics:}
We measure the performance of each agent in terms of six widely-adopted metrics inspired by the ANAC competition:
\begin{itemize}
    \item $U_{\it ind}^{total}$: The utility gained by an agent averaged over all the negotiations ($\uparrow$);
    \item $U_{\it ind}^s$: The utility gained by an agent averaged over all the \textit{successful} negotiations ($\uparrow$);    
    \item $U_{\it soc}$: The utility gained by both negotiating agents averaged over all successful negotiations ($\uparrow$);
    \item $P_{\it avg}$: Average minimal distance of agreements from the Pareto Frontier ($\downarrow$).
    \item $R_{\it avg}$: Average number of rounds before reaching the agreement ($\downarrow$);
    \item $S_{\%}$: Proportion of successful negotiations ($\uparrow$).
\end{itemize}
The first and second measures represent \textit{individual efficiency} of an outcome, whereas the third and fourth correspond to the \textit{social efficiency} of agreements.  

\paragraph{Experimental settings:}  \textit{ANESIA} is evaluated against state-of-the-art strategies that participated in ANAC'17, '18, and '19, and designed by different research groups independently. Each agent has no information about another agent's strategies beforehand. Details of all these strategies are available in~\cite{aydougan2018anac, jonker2020anac, aydougan2020challenges}.
We assume incomplete information about user preferences, given in the form of $B$ randomly-chosen partially-ordered bids. 
We evaluate \textit{ANESIA} on 8 negotiation domains which are different from each other in terms of size and opposition~\cite{baarslag2013evaluating} to ensure good negotiation characteristics and to reduce any biases. The domain size refers to the number of issues, whereas opposition\footnote{The value of opposition reflects the competitiveness between parties in the domain. Strong opposition means a gain of one party is at the loss of the other, whereas, weak opposition means that both parties either lose or gain simultaneously~\cite{baarslag2013evaluating}.}
refers to
the minimum distance from all possible outcomes to the point representing complete satisfaction of both negotiation parties (1,1). 
For our experiments, we choose readily-available 3 small-sized, 2 medium-sized, and 3 large-sized domains. Out of these domains, 2 are with high, 3 with medium and 3 with low opposition (see~\cite{williams2014overview} for more details). 
For each configuration, each agent plays both roles in the negotiation 
to compensate for any utility differences in the preference profiles. We call \textit{user profile} the agent's role along with the user's preferences. We set two user preferences uncertainties for each role: $|B| = 5\% |\Omega|$ and $|B|=10\% |\Omega|$.
Also, we set the $u_{res}$ and $d_D$ to their respective default values, whereas the deadline is set to 60s, normalized in $[0,1]$ (known to both negotiating parties in advance). 

Regarding the optimization algorithms, for FA (hypotheses A and C), we choose a population size of $20$ and $200$ generations for user model estimation and learning of strategy template parameters. We also set the maximum attractiveness value to $1.0$ and absorption coefficient to $0.01$. For NSGA-II (hypothesis B), we choose the population size of $2\%\times|\Omega|$, $2$ generations and mutation count of $0.1$.
With these hyperparameters, on our machine\footnote{CPU: 8 Cores, 2.10GHz; RAM: 32GB}
the run-time of NSGA-II never exceeded the given timeout of 10s for deciding an action at each turn, while being able to retrieve empirically good solutions. 
\vspace{-0.5mm}
\subsection{Empirical Evaluation}
\vspace{-2mm}
\subsubsection{Hypothesis A: User Modelling} 
We used two measures to determine the difference between 
$\estUsermodel$ and $\realUsermodel$~\cite{wachowicz2019tell}:
First, \textit{Ordinal accuracy (OA)} 
measures the proportion of bids put by $\widehat{U}$ in the correct rank order (i.e., as defined by the true user model), where an OA value of 1 implies a 100\% correct ranking. 
Second, to capture the scale of cardinal errors, \textit{Cardinal Inaccuracy (CI)} measures the differences in ratings assigned in the estimated and true user models for all the elements in domain $D$. 
We produced results in 8 domains and two profiles (5\% and 10\% of total possible bids) which are averaged over 10 simulations as shown in Table~\ref{table:usermodel}. All the values of OA ($\uparrow$) and CI ($\downarrow$), in each domain, for both the user profiles, are $\geq$ 0.67 and $\leq$ 0.90 respectively, 
which is quite accurate given the uncertainty and the fact that the CI value $\propto$ $|D|$.

\subsubsection{Hypothesis B: Pareto-Optimal Bids} 
We used a popular metric called  Inverted Generational Distance (IGD)~\cite{cheng2012performance} to compare the Pareto Fronts found by the NSGA-II and the ground truth (found via brute force).
Small IGD values suggest good convergence of solutions to the Pareto Front and their good distribution over the entire Pareto Front. 
Table~\ref{table:paretoIGD} demonstrates the potential of NSGA-II\footnote{Population size $=$ $0.02\times|\Omega|$; Number of generations $=$ 2.} for generating the Pareto-optimal bids as well as the closeness of true utility models. 
\begin{table*}[h]
    \centering
    \resizebox{.80\textwidth}{!}{
    \begin{tabular}{p{0.16\textwidth} p{0.16\textwidth}p{0.12\textwidth}
    p{0.12\textwidth}p{0.1\textwidth}
    p{0.1\textwidth}p{0.06\textwidth}}
     \multicolumn{7}{c}{(A) Performance Analysis of fully-fledged ANESIA}\\
    \hline
      Agent &
      $R_{\it avg} (\downarrow)$  &  
      $P_{\it avg} (\downarrow)$  &  
      $U_{\it soc} (\uparrow)$  &  
      $U_{\it ind}^{total} (\uparrow)$  &  
      $U_{\it ind}^s (\uparrow)$  &
      $S_{\%} (\uparrow)$ \\  
      \hline

    \textit{ANESIA} & \textcolor{blue}{626.24 $\pm$ 432.45}  & \textcolor{blue}{0.17 $\pm$ 0.29} & \textcolor{purple}{\textbf{1.51 $\pm$ 0.47}} & 0.66 $\pm$ 0.25 & \textcolor{purple}{\textbf{0.95 $\pm$ 0.06}} & 0.51 \\
     \hline
     \hline
    AgentGP $\bullet$ & 1366.95 $\pm$ 1378.19 & 0.30 $\pm$ 0.29 & 0.82 $\pm$ 0.58 & 0.66 $\pm$ 0.22 & 0.88 $\pm$ 0.09 & 0.58 \\
    FSEGA2019 $\bullet$& 1801.96 $\pm$ 1754.91 & \textcolor{blue}{0.17 $\pm$ 0.19} & 1.15 $\pm$ 0.37 & \textcolor{blue}{0.74 $\pm$ 0.18} & 0.82 $\pm$ 0.11 & 0.83 \\
     \hline
    \hline
    AgentHerb $\diamond$ & \textcolor{purple}{\textbf{32.30 $\pm$ 50.53}} & \textcolor{purple}{\textbf{0.01 $\pm$ 0.03}}& 1.41 $\pm$ 0.10 & 0.45 $\pm$ 0.13 & 0.45 $\pm$ 0.13 & \textcolor{purple}{\textbf{1.00}} \\
    Agent33 $\diamond$ & 4044.47 $\pm$ 4095.57 & 0.07 $\pm$ 0.16 & 1.31 $\pm$ 0.32 & 0.62 $\pm$ 0.15 & 0.64 $\pm$ 0.14 & 0.93 \\
    Sontag $\diamond$ & 5129.47 $\pm$ 5855.11 & 0.10 $\pm$ 0.18 & 1.22 $\pm$ 0.37 & 0.73 $\pm$ 0.17 & 0.79 $\pm$ 0.11 & 0.86 \\
    AgreeableAgent $\diamond$& 5003.98 $\pm$ 5216.68 & 0.11 $\pm$ 0.23 & 0.9 $\pm$ 0.4 & 0.67 $\pm$ 0.20 & 0.73 $\pm$ 0.13 & 0.73 \\
    PonpokoAgent $\star$ & 6609.73 $\pm$ 6611.18 & 0.18 $\pm$ 0.26 & 1.01 $\pm$ 0.49 & \textcolor{purple}{\textbf{0.76 $\pm$ 0.22}} & 0.89 $\pm$ 0.07 & 0.74 \\
    ParsCat2 $\star$& 4971.43 $\pm$ 4911.8 & 0.11 $\pm$ 0.22 & 1.16 $\pm$ 0.42 & 0.75 $\pm$ 0.19 & 0.81 $\pm$ 0.12 & 0.85 \\
    \hline
    \\[-5pt]
        \multicolumn{7}{c}{(B) Performance Analysis of ANESIA-DRL (Ablation study 1)}\\

    \hline
      Agent &
      $R_{\it avg} (\downarrow)$  &  
      $P_{\it avg} (\downarrow)$  &  
      $U_{\it soc} (\uparrow)$  &  
      $U_{\it ind}^{total} (\uparrow)$  &  
      $U_{\it ind}^s (\uparrow)$  &
      $S_{\%} (\uparrow)$ \\   
      \hline

     ANESIA-DRL & 480.43 $\pm$ 418.81 & \textcolor{blue}{0.10 $\pm$ 0.27} & \textcolor{purple}{\textbf{1.34 $\pm$ 0.51}} & 0.65 $\pm$ 0.22 & \textcolor{purple}{\textbf{0.87 $\pm$ 0.14}} & 0.56 \\
     \hline
     \hline
    AgentGP $\bullet$ & \textcolor{blue}{296.58 $\pm$ 340.15} & 0.24 $\pm$ 0.25 & 0.95 $\pm$ 0.48 & 0.61 $\pm$ 0.20 & 0.69 $\pm$ 0.15 & 0.75 \\
    FSEGA2019 $\bullet$ & 1488.63 $\pm$ 1924.03 & 0.22 $\pm$ 0.20 & 0.97 $\pm$ 0.39 & \textcolor{blue}{0.67 $\pm$ 0.20} & 0.79 $\pm$ 0.12 & \textcolor{blue}{0.77} \\
    \hline
    \hline
    AgentHerb $\diamond$ & \textcolor{purple}{\textbf{38.25 $\pm$ 79.16}} & \textcolor{purple}{\textbf{0.09 $\pm$ 0.09}} & 1.33 $\pm$ 0.15 & 0.48 $\pm$ 0.16 & 0.48 $\pm$ 0.16 & \textcolor{purple}{\textbf{0.99}} \\
    Agent33 $\diamond$& 2799.6 $\pm$ 3702.85 & 0.13 $\pm$ 0.15 & 1.22 $\pm$ 0.29 & 0.58 $\pm$ 0.15 & 0.59 $\pm$ 0.15 & 0.94 \\
    Sontag $\diamond$ & 3490.01 $\pm$ 4909.81 & 0.17 $\pm$ 0.19 & 1.09 $\pm$ 0.37 & 0.69 $\pm$ 0.18 & 0.76 $\pm$ 0.13 & 0.85 \\
    AgreeableAgent $\diamond$ & 3585.94 $\pm$ 4474.06 & 0.18 $\pm$ 0.21 & 0.78 $\pm$ 0.38 & 0.61 $\pm$ 0.19 & 0.68 $\pm$ 0.14 & 0.69 \\
    PonpokoAgent $\star$ & 4470.86 $\pm$ 5482.31 & 0.25 $\pm$ 0.24 & 0.90 $\pm$ 0.45 & \textcolor{purple}{\textbf{0.71 $\pm$ 0.20}} & 0.85 $\pm$ 0.09 & 0.72 \\
    ParsCat2 $\star$ & 3430.55 $\pm$ 4204.93 & 0.17 $\pm$ 0.20 & 1.07 $\pm$ 0.39 & 0.69 $\pm$ 0.18 & 0.73 $\pm$ 0.16 & 0.85 \\
        \hline
\\[-5pt]
    \multicolumn{7}{c}{(C) Performance analysis of Random ANESIA (Ablation study 2)}\\

    \hline
      Agent &
      $R_{\it avg} (\downarrow)$  &  
      $P_{\it avg} (\downarrow)$  &  
      $U_{\it soc} (\uparrow)$  &  
     $U_{\it ind}^{total} (\uparrow)$  &  
      $U_{\it ind}^s (\uparrow)$  &
      $S_{\%} (\uparrow)$ \\  
      \hline
   
     ANESIA-rand & 351.77 $\pm$ 343.24 & \textcolor{blue}{0.17 $\pm$ 0.21} & \textcolor{blue}{1.28 $\pm$ 0.43}   & 0.67 $\pm$ 0.22 & 0.74 $\pm$ 0.16 & 0.80 \\
    \hline
    \hline
    AgentGP $\bullet$ & \textcolor{blue}{258.12 $\pm$ 300.53} & 0.22 $\pm$ 0.24 & 1.0 $\pm$ 0.46 & 0.63 $\pm$ 0.19 & 0.70 $\pm$ 0.14 & 0.79 \\
    FSEGA2019 $\bullet$ & 1002.3 $\pm$ 1349.64 & 0.20 $\pm$ 0.19 & 1.02 $\pm$ 0.35 & \textcolor{blue}{0.68 $\pm$ 0.2} & \textcolor{blue}{0.79 $\pm$ 0.12} & \textcolor{blue}{0.81} \\
    \hline
    \hline
    AgentHerb $\diamond$ & \textcolor{purple}{\textbf{44.87 $\pm$ 80.16}} & \textcolor{purple}{\textbf{0.09 $\pm$ 0.09}} & \textcolor{purple}{\textbf{1.33 $\pm$ 0.15}} & 0.48 $\pm$ 0.15 & 0.48 $\pm$ 0.15 & \textcolor{purple}{\textbf{0.99}}\\
    Agent33 $\diamond$ & 1592.20 $\pm$ 2298.71 & 0.11 $\pm$ 0.12 & 1.26 $\pm$ 0.24 & 0.58 $\pm$ 0.15 & 0.58 $\pm$ 0.15 & 0.97 \\
    Sontag $\diamond$ & 2276.96 $\pm$ 3366.20 & 0.15 $\pm$ 0.17 & 1.13 $\pm$ 0.34 & 0.70 $\pm$ 0.17 & 0.75 $\pm$ 0.13 & 0.88 \\
    AgreeableAgent $\diamond$  & 2472.87 $\pm$ 3145.20 & 0.14 $\pm$ 0.17 & 0.86 $\pm$ 0.32 & 0.62 $\pm$ 0.18 & 0.67 $\pm$ 0.14 & 0.75 \\
    PonpokoAgent $\star$ & 2825.54 $\pm$ 3745.98 & 0.22 $\pm$ 0.23 & 0.96 $\pm$ 0.41 & \textcolor{purple}{\textbf{0.73 $\pm$ 0.20}}& \textcolor{purple}{\textbf{0.84 $\pm$ 0.09}} & 0.78 \\
    ParsCat2 $\star$ & 1992.60 $\pm$ 2674.90 & 0.13 $\pm$ 0.15 & 1.07 $\pm$ 0.30 & 0.66 $\pm$ 0.17 & 0.69 $\pm$ 0.14 & 0.86 \\
        \hline
    \end{tabular}
    }
    \caption{Performance comparison of \textit{ANESIA} against other ANAC winning strategies averaged over all the 8 domains and two uncertain preference profiles in each domain. See Appendix for the separate results for each domain. ANAC'19 agents ($\bullet$) have uncertain user
    preferences, and no learning capabilities. ANAC'17 ($\star$) and ANAC'18 ($\diamond$) agents can learn from experience and are given real user preferences. In {\textcolor{blue}{blue}} are the best among ANESIA and ANAC'19 agents. In {\textcolor{purple}{purple}}, the overall best.}
    \label{table:hypD}
    \vspace{-4mm}
\end{table*}

\subsubsection{Hypothesis C: \textit{ANESIA} outperforms ``teacher'' strategies}
We performed a total of $1440$ negotiation sessions\footnote{$n \times (n-1)/2 \times x \times y \times z \times w = 1440$ where $n = 4$, number of agents in a tournament; $x=2$, because agents play both sides; $y=3$, number of domains; $z=20$, because each tournament is repeated 20 times; $w=2$, number of profiles in terms of B.}
to evaluate the performance of \textit{ANESIA} against the three ``teacher'' strategies (AgentGG, KakeSoba and SAGA) in three domains (Laptop, Holiday, and Party) for two different profiles ($|B| = {10,20}$). These strategies were used to collect the dataset in the same domains for supervised training before the DRL process begins. Table~\ref{table:teacher} demonstrates the average results over all the domains and profiles for each agent. Clearly,  \textit{ANESIA} outperforms the ``teacher'' strategies in terms of $U_{\it ind}^{s}$ (i.e., individual efficiency), $U_{\it soc}$, and $P_{avg}$ (i.e., social efficiency). 

\subsubsection{Hypothesis D: Adaptive Behaviour of \textit{ANESIA} agent}
We further evaluated the performance of \textit{ANESIA} on agents (from ANAC'17, ANAC'18 and ANAC'19) unseen during training. 
For this, we performed a total of $23040$ negotiation sessions\footnote{$n \times (n-1)/2 \times x \times y \times z \times w = 23040$ where $n = 9$; $x=2$; $y=8$; $z=20$; and $w=2$.}.
Results in Table \ref{table:hypD}(A) are averaged over all domains and profiles, and  demonstrate that \textit{ANESIA} learns to make the optimal choice of tactics to be used at run time and outperforms the other $8$ strategies
in terms of $U_{\it ind}^{s}$ and $U_{\it soc}$. 

\paragraph{Ablation Study 1: } We evaluated the ANESIA-DRL performance, i.e., an \textit{ANESIA} agent that does not use templates to learn optimal combinations of tactics, but uses only one acceptance tactic, given by the dynamic DRL-based threshold utility $\bar{u}_t$ (and the Boulware and Pareto-optimal tactics for bidding) for the same negotiation settings of Hypothesis D. 
We observe from Table \ref{table:hypD}(B) that \textit{ANESIA-DRL} outperforms the other strategies in terms of $U_{\it ind}^{s}$ and $U_{\it soc}$.

\paragraph{Ablation Study 2: } We evaluated the ANESIA-rand performance, i.e., an \textit{ANESIA} agent which starts from a random DRL policy, i.e., without any offline pre-training of the adaptive utility threshold $\bar{u}_t$, for the same negotiation settings 
of Hypothesis D. 
From Table \ref{table:hypD}(B), we observe in ANESIA-rand some degradation of the utility metrics compared to the fully-fledged \textit{ANESIA} and ANESIA-DRL, even though it remains equally competitive w.r.t. the ANAC'17,18 agents and outperforms the ANAC'19 agents in $P_{avg}$, $U_{\it ind}^{s}$ and $U_{soc}$. This is not unexpected because, with a poorly informed (random) target utility tactic, the agent tends to
accept offers with little pay-off without negotiating for more rounds. 

\begin{figure}
    \centering
    \includegraphics[scale=0.95]{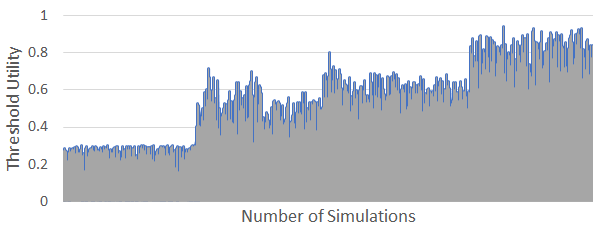}
    \caption{Increase in Dynamic Threshold Utility using DRL}
    \label{graph:threshold}
    \vspace{-4mm}
\end{figure}

From the Table \ref{table:hypD}(B) results, we conclude that the combination of both features (strategy templates and pre-training of the DRL model) are beneficial, even though these two features perform well in isolation too.  During the experiments, we observed that \textit{ANESIA} becomes picky by setting the dynamic threshold utility higher and hence learns to focus on getting the maximum utility from the end agreement. This leads to low success rate as compared to other agents. Figure~\ref{graph:threshold} shows an example of such threshold utility increase over time in one of the domains (Grocery) against a set of unknown opponents (Kakesoba and SAGA).
We note that AgentHerb is the best in terms of $P_{avg}$, which is not surprising because this is one of the agents that know the true user model. This is clearly an unfair advantage over the agents, like \textit{ANESIA}, that do not have this information.
That said, ANESIA attains the best $P_{avg}$ among ANAC'19 agents (unaware of the true user model) and the second best lowest $P_{avg}$ among ANAC'17 and ANAC'18 agents (aware of the true user model). 
Even though they have an unfair advantage in knowing the true user model, we consider ANAC'17 and ANAC'18 agents since, like our approach, they enable learning from past negotiations. 

To this end, we note that \textit{ANESIA} uses prior negotiation data from AgentGG to pre-train the DRL-based utility threshold and adjust the selection of tactics from the templates.
The effectiveness of our approach is demonstrated by the fact that \textit{ANESIA} outperforms the same agents it was trained on (see Hypothesis C), but, crucially, does so also on domains and opponents unseen during training. 
We further stress that the obtained performance metrics are affected only in part by an adequate pre-training of the strategies: the quality of the estimated user and opponent models -- derived without any prior training data from other agents -- plays an important role too. 
The results in Tables~\ref{table:hypD} (A to C)
evidence that our agent consistently outperforms its opponents in terms of individual and social efficiency, demonstrating  that \textit{ANESIA} can learn to adapt at run-time to different negotiation settings and against different unknown opponents.
\section{Conclusions}\label{conclusions}
\textit{ANESIA} is a novel agent model encapsulating different types of learning to support negotiation over multiple issues and under user preference uncertainty. An \textit{ANESIA} agent uses stochastic search based on FA for user modelling and combines NSGA-II and TOPSIS for generating Pareto bids during negotiation. 
It further exploits strategy templates to learn the best combination of acceptance and bidding tactics at any negotiation time, and among its tactics, it uses an adaptive target threshold utility learned using the DDPG algorithm. 
We have empirically evaluated the performance of \textit{ANESIA} against the winning agent strategies of ANAC'17, '18 and '19 competitions in different settings,  showing that our agent both outperforms opponents known at training time and can effectively transfer its knowledge to environments with previously unseen opponent agents and domains.

An open problem worth pursuing in the future is how to update the synthesized template strategy dynamically.

\newpage
\bibliography{ref.bib}
 \begin{figure*}
 \centering
    \includegraphics[scale=0.5]{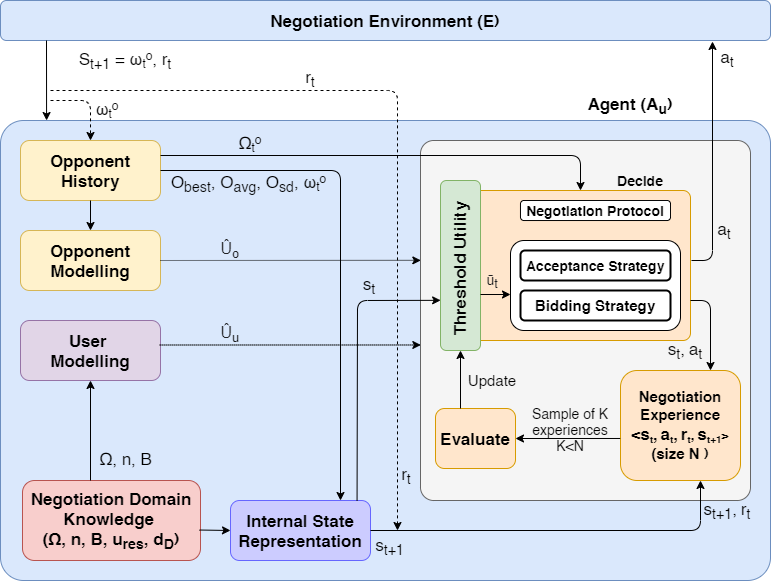}
    \caption{Interaction between the components of \textit{ANESIA}}
    \label{fig:anesia}
\end{figure*}
\newpage
\begin{table*}
    \centering
    \resizebox{.98\textwidth}{!}{
\begin{tabular}
    {p{0.07\textwidth} p{0.1\textwidth}
    p{0.1\textwidth}
    p{0.1\textwidth}p{0.1\textwidth}
    p{0.1\textwidth}p{0.1\textwidth}
    p{0.1\textwidth}p{0.1\textwidth}}
    \hline
    Metric & \multicolumn{2}{c}{ANESIA} & \multicolumn{2}{c}{AgentGG} & \multicolumn{2}{c}{KakeSoba} & 
    \multicolumn{2}{c}{SAGA}  \\
    \hline
    & $|B| = 5$ & $|B| = 10$   
    & $|B| = 5$ & $|B| = 10$  
    & $|B| = 5$ & $|B| = 10$
    & $|B| = 5$ & $|B| = 10$ \\
    \hline
    \multicolumn{9}{c}{\textbf{Holiday Domain}} \\
    \hline
    
    $R_{\it avg}$  &  476.81 $\pm$ 847.48 & 259.35 $\pm$ 453.3 &  1224.04  $\pm$ 1770.48 & 532.77  $\pm$ 841.16 & 1066.18 $\pm$ 1710.8	 & 414.51 $\pm$ 719.58 & 32.3 $\pm$ 83.79 & 238.18 $\pm$ 783.88\\
     
     $P_{\it avg}$  &  \textbf{0.15 $\pm$ 0.57} & \textbf{0.17 $\pm$ 0.26} & 0.4  $\pm$ 0.46 & 0.35  $\pm$ 0.44 & 0.28 $\pm$ 0.44 & 0.17 $\pm$ 0.34 & 0.23 $\pm$ 0.37 & 0.31 $\pm$ 0.42 \\
     
      $U_{\it soc}$  & \textbf{1.79 $\pm$ 0.81} & \textbf{1.72 $\pm$ 0.78} & 1.25  $\pm$0.66 & 1.33  $\pm$ 0.62 & 1.41 $\pm$ 0.63 & 1.57 $\pm$ 0.49 & 1.49 $\pm$ 0.53 & 1.39 $\pm$ 0.59\\
    
    $U_{\it ind}$  & 0.46 $\pm$ 0.42 & 0.51 $\pm$ 0.39 & 0.67  $\pm$0.37 & 0.73  $\pm$0.34 & 0.73 $\pm$ 0.34 & 0.8 $\pm$ 0.26	& 0.65 $\pm$ 0.24 & 0.61 $\pm$ 0.29\\
    
    $U_{\it ind}^s$  & \textbf{0.88 $\pm$ 0.11} & \textbf{0.88 $\pm$ 0.01} & 0.86  $\pm$0.13 & 0.88  $\pm$0.07 & 0.87 $\pm$ 0.13 & 0.87 $\pm$ 0.11 & 0.73 $\pm$ 0.1	 & 0.71 $\pm$ 0.17\\
\hline
    \end{tabular}
    }
    \caption{Performance Comparison of \textit{ANESIA} with ``teacher'' strategies. }
    \label{table:teacher1}
\end{table*}

\begin{table*}
    \centering
    \resizebox{.98\textwidth}{!}{
    \begin{tabular}{p{0.18\textwidth} p{0.18\textwidth}
    p{0.12\textwidth}p{0.12\textwidth}
    p{0.12\textwidth}p{0.12\textwidth} 
    p{0.1\textwidth}}
    \hline
    \multicolumn{7}{c}{Airport Domain B = 5\% of $\Omega$}\\
    \hline
      Agent &
      $R_{\it avg} (\downarrow)$  &  
      $P_{\it avg} (\downarrow)$  &  
      $U_{\it soc} (\uparrow)$  &  
      $U_{\it ind}^{total}  (\uparrow)$  &  
      $U_{\it ind}^s  (\uparrow)$  &
       $S_{\%} (\uparrow)$   \\  
      \hline
    ANESIA &  502.96 $\pm$ 350.71 & 0.06 $\pm$ 0.53 & \textbf{1.76 $\pm$ 0.79} & 0.69 $\pm$ 0.21 & \textbf{0.92 $\pm$ 0.11} & 0.48 \\
    \hline 
    \hline
    AgentGP $\bullet$ & 774.69 $\pm$ 562.07 & 0.59 $\pm$ 0.53 & 0.72 $\pm$ 0.79 & 0.69 $\pm$ 0.21 & 0.92 $\pm$ 0.06 & 0.45 \\
    FSEGA2019 $\bullet$ & 1023.41 $\pm$ 1450.49 & 0.07 $\pm$ 0.26 & 1.49 $\pm$ 0.38 & 0.77 $\pm$ 0.14 & 0.78 $\pm$ 0.13 & 0.94 \\
    \hline
    \hline
    AgentHerb $\diamond$  & \textbf{10.58 $\pm$ 3.59} & \textbf{0.0 $\pm$ 0.01} & 1.55 $\pm$ 0.11 & 0.59 $\pm$ 0.14 & 0.59 $\pm$ 0.14 & 1.00 \\
    Agent33 $\diamond$  & 1556.13 $\pm$ 2104.58 & 0.19 $\pm$ 0.41 & 1.3 $\pm$ 0.6 & 0.72 $\pm$ 0.13 & 0.76 $\pm$ 0.09 & 0.82 \\
    Sontag $\diamond$  & 1285.67 $\pm$ 1931.73 & 0.08 $\pm$ 0.27 & 1.47 $\pm$ 0.41 & 0.73 $\pm$ 0.12 & 0.75 $\pm$ 0.11 & 0.93 \\
\hline
    \end{tabular}
    }
    \caption{Performance of fully-fledged \textit{ANESIA} in the domain AIRPORT (1440 $\times 2$ profiles = 2880 simulations)}
    \label{table:AIRPORT}
\end{table*}

\begin{table*}
    \centering
    \resizebox{.98\textwidth}{!}{
     \begin{tabular}{p{0.18\textwidth} p{0.18\textwidth}
    p{0.12\textwidth}p{0.12\textwidth}
    p{0.12\textwidth}p{0.12\textwidth} 
    p{0.1\textwidth}}
    \hline
    \multicolumn{7}{c}{Camera Domain B = 5\% of $\Omega$}\\
    \hline
      Agent &
      $R_{\it avg} (\downarrow)$  &  
      $P_{\it avg} (\downarrow)$  &  
      $U_{\it soc} (\uparrow)$  &  
      $U_{\it ind}^{total}  (\uparrow)$  &  
      $U_{\it ind}^s  (\uparrow)$  &
       $S_{\%} (\uparrow)$   \\  
      \hline
    ANESIA  & 387.64 $\pm$ 264.96 & 0.02 $\pm$ 0.37 & \textbf{1.78 $\pm$ 0.66} & 0.74 $\pm$ 0.21 & \textbf{0.91 $\pm$ 0.08} & 0.59 \\
    \hline
    \hline
    AgentGP $\bullet$ & 615.41 $\pm$ 511.43 & 0.36 $\pm$ 0.37 & 0.78 $\pm$ 0.73 & 0.7 $\pm$ 0.2 & 0.87 $\pm$ 0.11 & 0.54 \\
    FSEGA2019 $\bullet$ & 58.92 $\pm$ 33.98 & 0.09 $\pm$ 0.23 & 1.28 $\pm$ 0.45 & 0.82 $\pm$ 0.12 & 0.86 $\pm$ 0.05 & 0.89 \\
    \hline
    \hline
     AgentHerb $\diamond$  & \textbf{7.48 $\pm$ 3.18} & \textbf{0.01 $\pm$ 0.02} & 1.43 $\pm$ 0.13 & 0.47 $\pm$ 0.15 & 0.47 $\pm$ 0.15 & \textbf{1.00} \\
    Agent33 $\diamond$  & 589.89 $\pm$ 802.57 & 0.01 $\pm$ 0.08 & 1.46 $\pm$ 0.21 & 0.71 $\pm$ 0.14 & 0.71 $\pm$ 0.14 & 0.99 \\
    Sontag $\diamond$ &  779.55 $\pm$ 1056.75 & 0.06 $\pm$ 0.2 & 1.37 $\pm$ 0.41 & 0.79 $\pm$ 0.12 & 0.81 $\pm$ 0.09 & 0.92 \\
\hline
    \end{tabular}
    }
    \caption{Performance of fully-fledged \textit{ANESIA} in the domain Camera (1440 $\times 2$ profiles = 2880 simulations)}
    \label{table:CAMERA}
\end{table*}

\begin{table*}
    \centering
    \resizebox{.98\textwidth}{!}{
    \begin{tabular}{p{0.18\textwidth} p{0.20\textwidth}
    p{0.12\textwidth}p{0.12\textwidth}
    p{0.12\textwidth}p{0.12\textwidth} 
    p{0.1\textwidth}}
    \hline
    \multicolumn{7}{c}{Energy Domain B = 5\% of $\Omega$}\\
    \hline
      Agent &
      $R_{\it avg} (\downarrow)$  &  
      $P_{\it avg} (\downarrow)$  &  
      $U_{\it soc} (\uparrow)$  &  
      $U_{\it ind}^{total}  (\uparrow)$  &  
      $U_{\it ind}^s  (\uparrow)$  &
       $S_{\%} (\uparrow)$   \\  
      \hline
     ANESIA & 1885.12 $\pm$ 1220.17 & 0.59 $\pm$ 0.24 & \textbf{1.15 $\pm$ 0.36} & 0.57 $\pm$ 0.17 & \textbf{0.97 $\pm$ 0.04} & 0.15 \\
     \hline
     \hline
    AgentGP $\bullet$ & 776.26 $\pm$ 1337.29 & 0.36 $\pm$ 0.34 & 0.58 $\pm$ 0.59 & \textbf{0.71 $\pm$ 0.21} & 0.92 $\pm$ 0.02 & 0.49 \\
    FSEGA2019 $\bullet$ &  227.55 $\pm$ 94.34 & 0.53 $\pm$ 0.29 & 0.26 $\pm$ 0.46 & 0.6 $\pm$ 0.18 & 0.91 $\pm$ 0.06 & 0.24 \\
    \hline
    \hline
     AgentHerb $\diamond$  & \textbf{209.5 $\pm$ 390.96} & \textbf{0.01 $\pm$ 0.03 }& 1.11 $\pm$ 0.13 & 0.21 $\pm$ 0.15 & 0.21 $\pm$ 0.15 & \textbf{1.00} \\
    Agent33 $\diamond$ & 9736.06 $\pm$ 11596.61 & 0.11 $\pm$ 0.24 & 0.98 $\pm$ 0.42 & 0.5 $\pm$ 0.16 & 0.5 $\pm$ 0.17 & 0.86 \\
    Sontag $\diamond$ & 16164.98 $\pm$ 19874.32 & 0.3 $\pm$ 0.33 & 0.65 $\pm$ 0.56 & 0.66 $\pm$ 0.17 & 0.78 $\pm$ 0.12 & 0.59 \\
\hline
    \end{tabular}
    }
    \caption{Performance of fully-fledged \textit{ANESIA} in the domain Energy (1440 $\times 2$ profiles = 2880 simulations)}
    \label{table:Energy}
\end{table*}

\begin{table*}
    \centering
    \resizebox{.98\textwidth}{!}{
     \begin{tabular}{p{0.18\textwidth} p{0.18\textwidth}
    p{0.12\textwidth}p{0.12\textwidth}
    p{0.12\textwidth}p{0.12\textwidth} 
    p{0.1\textwidth}}
    \hline
    \multicolumn{7}{c}{Grocery Domain B = 5\% of $\Omega$}\\
    \hline
      Agent &
      $R_{\it avg} (\downarrow)$  &  
      $P_{\it avg} (\downarrow)$  &  
      $U_{\it soc} (\uparrow)$  &  
      $U_{\it ind}^{total}  (\uparrow)$  &  
      $U_{\it ind}^s  (\uparrow)$  &
       $S_{\%} (\uparrow)$   \\  
      \hline
   
ANESIA & 358.62 $\pm$ 240.22 & 0.26 $\pm$ 0.37 & \textbf{1.57 $\pm$ 0.65} & 0.65 $\pm$ 0.43 &\textbf{ 0.93 $\pm$ 0.06} & 0.70 \\
\hline
\hline
AgentGP $\bullet$ & 507.2 $\pm$ 380.16 & 0.27 $\pm$ 0.37 & 1.0 $\pm$ 0.68 & 0.62 $\pm$ 0.42 & 0.90 $\pm$ 0.09 & 0.69 \\
FSEGA2019 $\bullet$ & 379.88 $\pm$ 415.61 & 0.01 $\pm$ 0.07 & 1.49 $\pm$ 0.17 & \textbf{0.84 $\pm$ 0.13} & 0.84 $\pm$ 0.11 & 0.99 \\
\hline
\hline
AgentHerb $\diamond$ & \textbf{5.54 $\pm$ 1.73} & 0.02 $\pm$ 0.05 & 1.53 $\pm$ 0.1 & 0.56 $\pm$ 0.12 & 0.56 $\pm$ 0.12 & \textbf{1.00} \\
Agent33 $\diamond$ & 300.19 $\pm$ 371.5 & 0.03 $\pm$ 0.1 & 1.51 $\pm$ 0.22 & 0.67 $\pm$ 0.16 & 0.68 $\pm$ 0.14 & 0.99 \\
Sontag $\diamond$ &  606.43 $\pm$ 837.78 & \textbf{0.01 $\pm$ 0.01} & 1.53 $\pm$ 0.12 & 0.82 $\pm$ 0.11 & 0.82 $\pm$ 0.11 & \textbf{1.00} \\
\hline
    \end{tabular}
    }
    \caption{Performance of fully-fledged \textit{ANESIA} in the domain Grocery (1440 $\times 2$ profiles = 2880 simulations)}
    \label{table:GROCERY}
\end{table*}

\begin{table*}
    \centering
    \resizebox{.98\textwidth}{!}{
    \begin{tabular}{p{0.18\textwidth} p{0.20\textwidth}
    p{0.12\textwidth}p{0.12\textwidth}
    p{0.12\textwidth}p{0.12\textwidth} 
    p{0.1\textwidth}}
    \hline
    \multicolumn{7}{c}{Fitness Domain B = 5\% of $\Omega$}\\
    \hline
      Agent &
      $R_{\it avg} (\downarrow)$  &  
      $P_{\it avg} (\downarrow)$  &  
      $U_{\it soc} (\uparrow)$  &  
      $U_{\it ind}^{total}  (\uparrow)$  &  
      $U_{\it ind}^s  (\uparrow)$  &
       $S_{\%} (\uparrow)$   \\  
      \hline
    
    ANESIA & 8.13 $\pm$ 1.19 & \textbf{0.0 $\pm$ 0.0} & \textbf{1.51 $\pm$ 0.01} &\textbf{ 1.0 $\pm$ 0.0} &\textbf{ 1.0 $\pm$ 0.0} & \textbf{1.00} \\
    \hline\hline
    AgentGP $\bullet$  & 5025.1 $\pm$ 4975.25 & 0.01 $\pm$ 0.01 & 1.49 $\pm$ 0.05 & 0.92 $\pm$ 0.02 & 0.92 $\pm$ 0.02 & \textbf{1.00 }\\
    FSEGA2019 $\bullet$ & 6462.11 $\pm$ 5295.59 &\textbf{ 0.0 $\pm$ 0.01} & 1.45 $\pm$ 0.07 & 0.81 $\pm$ 0.11 & 0.81 $\pm$ 0.11 & \textbf{1.00 }\\
    \hline\hline
     AgentHerb $\diamond$  & \textbf{7.25 $\pm$ 2.01} & 0.01 $\pm$ 0.02 & \textbf{1.51 $\pm$ 0.04} & 0.55 $\pm$ 0.07 & 0.55 $\pm$ 0.07 & \textbf{1.00} \\
    Agent33 $\diamond$ & 3483.88 $\pm$ 2350.29 & 0.01 $\pm$ 0.02 & 1.5 $\pm$ 0.04 & 0.7 $\pm$ 0.09 & 0.7 $\pm$ 0.09 &\textbf{ 1.00} \\
    Sontag $\diamond$  & 6117.57 $\pm$ 6137.66 & \textbf{0.0 $\pm$ 0.01} & 1.47 $\pm$ 0.07 & 0.8 $\pm$ 0.11 & 0.8 $\pm$ 0.11 & \textbf{1.00} \\
\hline
    \end{tabular}
    }
    \caption{Performance of fully-fledged \textit{ANESIA} in the domain Fitness (1440 $\times 2$ profiles = 2880 simulations)}
    \label{table:FITNESS}
\end{table*}

\begin{table*}
    \centering
    \resizebox{.98\textwidth}{!}{
     \begin{tabular}{p{0.18\textwidth} p{0.18\textwidth}
    p{0.12\textwidth}p{0.12\textwidth}
    p{0.12\textwidth}p{0.12\textwidth} 
    p{0.1\textwidth}}
    \hline
    \multicolumn{7}{c}{Flight Booking Domain B = 5\% of $\Omega$}\\
    \hline
      Agent &
      $R_{\it avg} (\downarrow)$  &  
      $P_{\it avg} (\downarrow)$  &  
      $U_{\it soc} (\uparrow)$  &  
      $U_{\it ind}^{total}  (\uparrow)$  &  
      $U_{\it ind}^s  (\uparrow)$  &
       $S_{\%} (\uparrow)$   \\  
      \hline
    ANESIA & 457.11 $\pm$ 307.19 & 0.04 $\pm$ 0.28 & \textbf{1.54 $\pm$ 0.63} & 0.66 $\pm$ 0.21 & \textbf{0.9 $\pm$ 0.16} & 0.43 \\
    \hline\hline
    AgentGP $\bullet$ & 664.13 $\pm$ 477.03 & 0.35 $\pm$ 0.28 & 0.56 $\pm$ 0.68 & 0.66 $\pm$ 0.2 & \textbf{0.9 $\pm$ 0.08} & 0.41 \\
    FSEGA2019 $\bullet$ & 1021.27 $\pm$ 1249.01 & 0.09 $\pm$ 0.2 & 1.14 $\pm$ 0.46 & 0.78 $\pm$ 0.15 & 0.82 $\pm$ 0.12 & 0.87 \\
    \hline\hline
     AgentHerb $\diamond$  & \textbf{7.7 $\pm$ 3.9} & \textbf{0.01 $\pm$ 0.05} & 1.37 $\pm$ 0.09 & 0.42 $\pm$ 0.13 & 0.42 $\pm$ 0.13 &\textbf{1.00} \\
    Agent33 $\diamond$  & 543.69 $\pm$ 758.12 & 0.05 $\pm$ 0.13 & 1.26 $\pm$ 0.34 & 0.52 $\pm$ 0.11 & 0.52 $\pm$ 0.11 & 0.95 \\
    Sontag $\diamond$  & 1041.66 $\pm$ 1365.84 & 0.1 $\pm$ 0.21 & 1.13 $\pm$ 0.51 & 0.75 $\pm$ 0.15 & 0.8 $\pm$ 0.11 & 0.84 \\
\hline
    \end{tabular}
    }
    \caption{Performance of fully-fledged \textit{ANESIA} in the domain Flight Booking (1440 $\times 2$ profiles = 2880 simulations)}
    \label{table:FLIGHT}
\end{table*}

\begin{table*}
    \centering
    \resizebox{.98\textwidth}{!}{
   \begin{tabular}{p{0.18\textwidth} p{0.18\textwidth}
    p{0.12\textwidth}p{0.12\textwidth}
    p{0.12\textwidth}p{0.12\textwidth} 
    p{0.1\textwidth}}
    \hline
    \multicolumn{7}{c}{Itex Domain B = 5\% of $\Omega$}\\
    \hline
      Agent &
      $R_{\it avg} (\downarrow)$  &  
      $P_{\it avg} (\downarrow)$  &  
      $U_{\it soc} (\uparrow)$  &  
      $U_{\it ind}^{total}  (\uparrow)$  &  
      $U_{\it ind}^s  (\uparrow)$  &
       $S_{\%} (\uparrow)$   \\  
      \hline
    
    ANESIA  & 564.37 $\pm$ 402.32 & 0.39 $\pm$ 0.15 & \textbf{1.26 $\pm$ 0.09} & 0.56 $\pm$ 0.17 & \textbf{1.0 $\pm$ 0.0} & 0.13 \\
    \hline\hline
    AgentGP $\bullet$ & 686.32 $\pm$ 481.19 & 0.32 $\pm$ 0.2 & 0.31 $\pm$ 0.5 & 0.6 $\pm$ 0.18 & 0.85 $\pm$ 0.14 & 0.29 \\
    FSEGA2019 $\bullet$  & 1833.21 $\pm$ 2581.66 & 0.14 $\pm$ 0.21 & 0.64 $\pm$ 0.46 & 0.67 $\pm$ 0.18 & 0.75 $\pm$ 0.17 & 0.68 \\
    \hline\hline
    AgentHerb $\diamond$ & \textbf{9.1 $\pm$ 2.82} & \textbf{0.0 $\pm$ 0.02} & 1.19 $\pm$ 0.04 & 0.23 $\pm$ 0.08 & 0.23 $\pm$ 0.08 & 1.00 \\
    Agent33 $\diamond$  & 1187.91 $\pm$ 2093.66 & 0.11 $\pm$ 0.19 & 0.8 $\pm$ 0.49 & 0.47 $\pm$ 0.16 & 0.46 $\pm$ 0.18 & 0.76 \\
    Sontag $\diamond$  & 1934.54 $\pm$ 2764.67 & 0.14 $\pm$ 0.2 & 0.65 $\pm$ 0.47 & \textbf{0.70 $\pm$ 0.17} & 0.79 $\pm$ 0.13 & 0.69 \\
\hline
    \end{tabular}
    }
    \caption{Performance of fully-fledged \textit{ANESIA} in the domain Itex (1440 $\times 2$ profiles = 2880 simulations)}
    \label{table:ITEX}
\end{table*}

\begin{table*}
    \centering
    \resizebox{.98\textwidth}{!}{
  \begin{tabular}{p{0.18\textwidth} p{0.18\textwidth}
    p{0.12\textwidth}p{0.12\textwidth}
    p{0.12\textwidth}p{0.12\textwidth} 
    p{0.1\textwidth}}
    \hline
    \multicolumn{7}{c}{Outfit Domain B = 5\% of $\Omega$}\\
    \hline
      Agent &
      $R_{\it avg} (\downarrow)$  &  
      $P_{\it avg} (\downarrow)$  &  
      $U_{\it soc} (\uparrow)$  &  
      $U_{\it ind}^{total}  (\uparrow)$  &  
      $U_{\it ind}^s  (\uparrow)$  &
       $S_{\%} (\uparrow)$   \\  
      \hline
    
    ANESIA & 516.7 $\pm$ 352.5 & 0.05 $\pm$ 0.4 &\textbf{ 1.57 $\pm$ 0.73} & 0.67 $\pm$ 0.37 & \textbf{0.99 $\pm$ 0.02} & 0.56 \\
    \hline\hline
    AgentGP $\bullet$ &  786.93 $\pm$ 580.41 & 0.39 $\pm$ 0.4 & 0.83 $\pm$ 0.79 & 0.61 $\pm$ 0.34 & 0.93 $\pm$ 0.07 & 0.53 \\
    FSEGA2019 $\bullet$  & 854.18 $\pm$ 890.84 & 0.05 $\pm$ 0.19 & 1.44 $\pm$ 0.39 & 0.79 $\pm$ 0.18 & 0.83 $\pm$ 0.11 & 0.94 \\
    \hline\hline
     AgentHerb $\diamond$  & \textbf{5.73 $\pm$ 3.37} &\textbf{ 0.01 $\pm$ 0.04} &\textbf{ 1.57 $\pm$ 0.16} & 0.60 $\pm$ 0.17 & 0.6 $\pm$ 0.17 & \textbf{1.00 }\\
    Agent33 $\diamond$  & 499.38 $\pm$ 763.55 & 0.04 $\pm$ 0.16 & 1.48 $\pm$ 0.37 & 0.68 $\pm$ 0.17 & 0.70 $\pm$ 0.15 & 0.96 \\
    Sontag $\diamond$  & 749.45 $\pm$ 868.18 & 0.05 $\pm$ 0.18 & 1.47 $\pm$ 0.38 & 0.78 $\pm$ 0.17 & 0.81 $\pm$ 0.12 & 0.94 \\
\hline
    \end{tabular}
    }
    \caption{Performance of fully-fledged \textit{ANESIA} in the domain Outfit (1440 $\times 2$ profiles = 2880 simulations)}
    \label{table:OUTFIT}
\end{table*}

\begin{table*}
    \centering
    \resizebox{.98\textwidth}{!}{
     \begin{tabular}{p{0.18\textwidth} p{0.20\textwidth}
    p{0.12\textwidth}p{0.12\textwidth}
    p{0.12\textwidth}p{0.12\textwidth} 
    p{0.1\textwidth}}
    \hline
    \multicolumn{7}{c}{Airport Domain B = 5\% of $\Omega$}\\
    \hline
      Agent &
      $R_{\it avg} (\downarrow)$  &  
      $P_{\it avg} (\downarrow)$  &  
      $U_{\it soc} (\uparrow)$  &  
      $U_{\it ind}^{total}  (\uparrow)$  &  
      $U_{\it ind}^s  (\uparrow)$  &
       $S_{\%} (\uparrow)$   \\  
      \hline
    
    ANESIA-DRL  & 414.88 $\pm$ 381.14 & \textbf{0.1 $\pm$ 0.5} & \textbf{1.43 $\pm$ 0.71} & 0.61 $\pm$ 0.16 & \textbf{0.86 $\pm$ 0.17} & 0.58 \\
    \hline\hline
    AgentGP $\bullet$ & 159.14 $\pm$ 137.0 & 0.08 $\pm$ 0.22 & \textbf{1.43 $\pm$ 0.32} & 0.64 $\pm$ 0.13 & 0.65 $\pm$ 0.13 & 0.96 \\
    FSEGA2019 $\bullet$ & 994.88 $\pm$ 1324.82 & \textbf{0.1 $\pm$ 0.24} & 1.41 $\pm$ 0.34 & 0.75 $\pm$ 0.17 & 0.75 $\pm$ 0.16 & 0.96 \\
    \hline\hline
    AgentHerb $\diamond$  & \textbf{9.9 $\pm$ 3.64} & 0.19 $\pm$ 0.22 & 1.28 $\pm$ 0.24 & 0.5 $\pm$ 0.23 & 0.5 $\pm$ 0.23 & \textbf{1.00 }\\
    Agent33 $\diamond$  & 971.03 $\pm$ 1467.42 & 0.25 $\pm$ 0.31 & 1.16 $\pm$ 0.44 & 0.55 $\pm$ 0.16 & 0.55 $\pm$ 0.17 & 0.93 \\
    Sontag $\diamond$  & 1101.3 $\pm$ 1448.19 & 0.14 $\pm$ 0.28 & 1.35 $\pm$ 0.41 & 0.72 $\pm$ 0.19 & 0.74 $\pm$ 0.18 & 0.94 \\
\hline
    \end{tabular}
    }
    \caption{Performance of ANESIA-DRL - Ablation Study1 - over domain AIRPORT (1440 $\times 2$ profiles = 2880 simulations)}
    \label{table:AIRPORT}
\end{table*}

\begin{table*}
    \centering
    \resizebox{.98\textwidth}{!}{
     \begin{tabular}{p{0.18\textwidth} p{0.20\textwidth}
    p{0.12\textwidth}p{0.12\textwidth}
    p{0.12\textwidth}p{0.12\textwidth} 
    p{0.1\textwidth}}
    \hline
    \multicolumn{7}{c}{Camera Domain B = 5\% of $\Omega$}\\
    \hline
      Agent &
      $R_{\it avg} (\downarrow)$  &  
      $P_{\it avg} (\downarrow)$  &  
      $U_{\it soc} (\uparrow)$  &  
      $U_{\it ind}^{total}  (\uparrow)$  &  
      $U_{\it ind}^s  (\uparrow)$  &
       $S_{\%} (\uparrow)$   \\  
      \hline
    
    ANESIA-DRL  & 340.58 $\pm$ 320.93 & \textbf{0.04 $\pm$ 0.32} & \textbf{1.47 $\pm$ 0.62} & 0.64 $\pm$ 0.14 & \textbf{0.86 $\pm$ 0.11} & 0.57 \\
    \hline\hline
    AgentGP $\bullet$ & 105.21 $\pm$ 268.21 & 0.17 $\pm$ 0.23 & 1.12 $\pm$ 0.45 & 0.65 $\pm$ 0.19 & 0.67 $\pm$ 0.19 & 0.88 \\
     FSEGA2019 $\bullet$ & 59.61 $\pm$ 38.96 & 0.19 $\pm$ 0.29 & 1.03 $\pm$ 0.55 & 0.76 $\pm$ 0.16 & 0.83 $\pm$ 0.1 & 0.80 \\
    \hline\hline
    AgentHerb $\diamond$  & \textbf{4.99 $\pm$ 2.94} & 0.06 $\pm$ 0.08 & 1.41 $\pm$ 0.2 & 0.53 $\pm$ 0.22 & 0.53 $\pm$ 0.22 & \textbf{1.00} \\
   
    Agent33 $\diamond$ & 422.69 $\pm$ 630.12 & 0.06 $\pm$ 0.07 & 1.41 $\pm$ 0.19 & 0.69 $\pm$ 0.18 & 0.69 $\pm$ 0.18 & \textbf{1.00 }\\
    Sontag $\diamond$  & 704.9 $\pm$ 974.92 & 0.07 $\pm$ 0.14 & 1.35 $\pm$ 0.29 & 0.74 $\pm$ 0.15 & 0.74 $\pm$ 0.15 & 0.97 \\
\hline
    \end{tabular}
    }
    \caption{Performance of ANESIA-DRL - Ablation Study1 - over domain Camera (1440 $\times 2$ profiles = 2880 simulations)}
    \label{table:CAMERA}
\end{table*}

\begin{table*}
    \centering
    \resizebox{.98\textwidth}{!}{
     \begin{tabular}{p{0.18\textwidth} p{0.20\textwidth}
    p{0.12\textwidth}p{0.12\textwidth}
    p{0.12\textwidth}p{0.12\textwidth} 
    p{0.1\textwidth}}
    \hline
    \multicolumn{7}{c}{Energy Domain B = 5\% of $\Omega$}\\
    \hline
      Agent &
      $R_{\it avg} (\downarrow)$  &  
      $P_{\it avg} (\downarrow)$  &  
      $U_{\it soc} (\uparrow)$  &  
      $U_{\it ind}^{total}  (\uparrow)$  &  
      $U_{\it ind}^s  (\uparrow)$  &
       $S_{\%} (\uparrow)$   \\  
      \hline
    ANESIA-DRL  & 1397.63 $\pm$ 1143.1 & 0.08 $\pm$ 0.28 & \textbf{1.09 $\pm$ 0.49} & 0.56 $\pm$ 0.12 & \textbf{0.89 $\pm$ 0.16} & 0.36 \\
    \hline\hline
    AgentGP $\bullet$ & 131.72 $\pm$ 319.48 & 0.29 $\pm$ 0.28 & 0.66 $\pm$ 0.48 & 0.5 $\pm$ 0.16 & 0.51 $\pm$ 0.19 & 0.68 \\
    FSEGA2019 $\bullet$ & 227.86 $\pm$ 126.06 & 0.57 $\pm$ 0.24 & 0.19 $\pm$ 0.39 & 0.57 $\pm$ 0.14 & 0.84 $\pm$ 0.09 & 0.20 \\
    \hline\hline
    AgentHerb $\diamond$ & \textbf{22.92 $\pm$ 14.14} & \textbf{0.07 $\pm$ 0.07} & 1.08 $\pm$ 0.08 & 0.24 $\pm$ 0.18 & 0.24 $\pm$ 0.18 & \textbf{1.00} \\
    Agent33 $\diamond$  & 10734.09 $\pm$ 15636.62 & 0.22 $\pm$ 0.24 & 0.81 $\pm$ 0.43 & 0.51 $\pm$ 0.1 & 0.52 $\pm$ 0.12 & 0.80 \\
    Sontag $\diamond$  & 12776.72 $\pm$ 18396.44 & 0.33 $\pm$ 0.31 & 0.57 $\pm$ 0.5 & \textbf{0.65 $\pm$ 0.17} & 0.76 $\pm$ 0.14 & 0.57 \\
\hline
    \end{tabular}
    }
    \caption{Performance of ANESIA-DRL - Ablation Study1 - over domain Energy (1440 $\times 2$ profiles = 2880 simulations)}
    \label{table:ENERGY}
\end{table*}

\begin{table*}
    \centering
    \resizebox{.98\textwidth}{!}{
    \begin{tabular}{p{0.18\textwidth} p{0.20\textwidth}
    p{0.12\textwidth}p{0.12\textwidth}
    p{0.12\textwidth}p{0.12\textwidth} 
    p{0.1\textwidth}}
    \hline
    \multicolumn{7}{c}{Grocery Domain B = 5\% of $\Omega$}\\
    \hline
      Agent &
      $R_{\it avg} (\downarrow)$  &  
      $P_{\it avg} (\downarrow)$  &  
      $U_{\it soc} (\uparrow)$  &  
      $U_{\it ind}^{total}  (\uparrow)$  &  
      $U_{\it ind}^s  (\uparrow)$  &
       $S_{\%} (\uparrow)$   \\  
      \hline
    
   ANESIA-DRL  & 277.41 $\pm$ 266.23 & 0.35 $\pm$ 0.3 & \textbf{1.24 $\pm$ 0.18} & 0.55 $\pm$ 0.35 & \textbf{0.76 $\pm$ 0.12} & 0.73 \\
   \hline\hline
    AgentGP $\bullet$  & 102.75 $\pm$ 249.25 & 0.24 $\pm$ 0.2 & 1.22 $\pm$ 0.41 & 0.68 $\pm$ 0.24 & 0.75 $\pm$ 0.13 & 0.91 \\
    FSEGA2019 $\bullet$  & 244.66 $\pm$ 356.63 & 0.25 $\pm$ 0.15 & 1.18 $\pm$ 0.32 & 0.62 $\pm$ 0.2 & 0.65 $\pm$ 0.13 & 0.95 \\
   \hline\hline
    AgentHerb $\diamond$ & \textbf{5.22 $\pm$ 2.17} & 0.26 $\pm$ 0.08 & 1.23 $\pm$ 0.15 & 0.51 $\pm$ 0.15 & 0.51 $\pm$ 0.15 & \textbf{1.00} \\
    Agent33 $\diamond$ & 90.96 $\pm$ 103.58 & 0.25 $\pm$ 0.08 & \textbf{1.24 $\pm$ 0.15} & 0.57 $\pm$ 0.17 & 0.57 $\pm$ 0.17 & 1.00 \\
    Sontag $\diamond$  & 421.19 $\pm$ 754.88 & 0.22 $\pm$ 0.09 & \textbf{1.24 $\pm$ 0.19} & 0.65 $\pm$ 0.15 & 0.65 $\pm$ 0.13 & 0.99 \\
\hline
    \end{tabular}
    }
    \caption{Performance of ANESIA-DRL - Ablation Study1 - over domain Grocery (1440 $\times 2$ profiles = 2880 simulations)}
    \label{table:AIRPORT}
\end{table*}

\begin{table*}
    \centering
    \resizebox{.98\textwidth}{!}{
     \begin{tabular}{p{0.18\textwidth} p{0.20\textwidth}
    p{0.12\textwidth}p{0.12\textwidth}
    p{0.12\textwidth}p{0.12\textwidth} 
    p{0.1\textwidth}}
    \hline
    \multicolumn{7}{c}{Fitness Domain B = 5\% of $\Omega$}\\
    \hline
      Agent &
      $R_{\it avg} (\downarrow)$  &  
      $P_{\it avg} (\downarrow)$  &  
      $U_{\it soc} (\uparrow)$  &  
      $U_{\it ind}^{total}  (\uparrow)$  &  
      $U_{\it ind}^s  (\uparrow)$  &
       $S_{\%} (\uparrow)$   \\  
      \hline
    
   ANESIA-DRL & \textbf{6.5 $\pm$ 2.6} & \textbf{0.05 $\pm$ 0.06} & \textbf{1.48 $\pm$ 0.07} & \textbf{0.87 $\pm$ 0.15} &\textbf{ 0.87 $\pm$ 0.15} & \textbf{1.00} \\
   \hline\hline
    AgentGP $\bullet$ & 507.39 $\pm$ 1234.05 & 0.1 $\pm$ 0.2 & 1.32 $\pm$ 0.37 & 0.67 $\pm$ 0.16 & 0.68 $\pm$ 0.16 & 0.93 \\

    FSEGA2019 $\bullet$ & 2142.49 $\pm$ 3180.88 & 0.09 $\pm$ 0.14 & 1.32 $\pm$ 0.25 & 0.77 $\pm$ 0.13 & 0.78 $\pm$ 0.12 & 0.97 \\
    \hline\hline

    AgentHerb $\diamond$ & 7.36 $\pm$ 3.51 &\textbf{ 0.05 $\pm$ 0.05} & 1.48 $\pm$ 0.05 & 0.58 $\pm$ 0.1 & 0.58 $\pm$ 0.1 &\textbf{ 1.00 }\\
    Agent33 $\diamond$ & 761.68 $\pm$ 1289.88 & 0.07 $\pm$ 0.05 & 1.44 $\pm$ 0.07 & 0.68 $\pm$ 0.12 & 0.68 $\pm$ 0.12 &\textbf{ 1.00} \\
    Sontag $\diamond$ & 2813.95 $\pm$ 5044.51 & 0.07 $\pm$ 0.07 & 1.36 $\pm$ 0.14 & 0.78 $\pm$ 0.13 & 0.79 $\pm$ 0.13 & \textbf{1.00} \\
\hline
    \end{tabular}
    }
    \caption{Performance of ANESIA-DRL - Ablation Study1 - over domain Fitness (1440 $\times 2$ profiles = 2880 simulations)}
    \label{table:AIRPORT}
\end{table*}

\begin{table*}
    \centering
    \resizebox{.98\textwidth}{!}{
    \begin{tabular}{p{0.18\textwidth} p{0.20\textwidth}
    p{0.12\textwidth}p{0.12\textwidth}
    p{0.12\textwidth}p{0.12\textwidth} 
    p{0.1\textwidth}}
    \hline
    \multicolumn{7}{c}{Flight Booking Domain B = 5\% of $\Omega$}\\
    \hline
      Agent &
      $R_{\it avg} (\downarrow)$  &  
      $P_{\it avg} (\downarrow)$  &  
      $U_{\it soc} (\uparrow)$  &  
      $U_{\it ind}^{total}  (\uparrow)$  &  
      $U_{\it ind}^s  (\uparrow)$  &
       $S_{\%} (\uparrow)$   \\  
      \hline
    
   ANESIA-DRL  & 636.1 $\pm$ 658.93 & 0.13 $\pm$ 0.25 & \textbf{1.36 $\pm$ 0.58} & 0.59 $\pm$ 0.18 & \textbf{0.98 $\pm$ 0.2} & 0.29 \\
   \hline\hline
    AgentGP $\bullet$ & 505.74 $\pm$ 291.15 & 0.41 $\pm$ 0.24 & 0.47 $\pm$ 0.61 & 0.55 $\pm$ 0.12 & 0.62 $\pm$ 0.16 & 0.39 \\
    FSEGA2019 $\bullet$ & 3733.36 $\pm$ 5491.65 & 0.47 $\pm$ 0.23 & 0.23 $\pm$ 0.48 & 0.6 $\pm$ 0.2 & 1.0 $\pm$ 0.0 & 0.20 \\
    \hline\hline
    AgentHerb $\diamond$ & \textbf{451.6 $\pm$ 1069.53} & \textbf{0.12 $\pm$ 0.19} & 1.18 $\pm$ 0.44 & 0.51 $\pm$ 0.16 & 0.51 $\pm$ 0.17 & \textbf{0.89} \\
    Agent33 $\diamond$  & 1446.79 $\pm$ 3331.37 & 0.17 $\pm$ 0.23 & 0.98 $\pm$ 0.52 & 0.51 $\pm$ 0.17 & 0.51 $\pm$ 0.19 & 0.81 \\
    Sontag $\diamond$ & 2994.05 $\pm$ 5633.23 & 0.36 $\pm$ 0.27 & 0.53 $\pm$ 0.65 & 0.67 $\pm$ 0.22 & 0.93 $\pm$ 0.09 & 0.41 \\
\hline
    \end{tabular}
    }
    \caption{Performance of ANESIA-DRL - Ablation Study1 - over domain Flight Booking (1440 $\times 2$ profiles = 2880 simulations)}
    \label{table:FLIGHT}
\end{table*}

\begin{table*}
    \centering
    \resizebox{.98\textwidth}{!}{
    \begin{tabular}{p{0.18\textwidth} p{0.20\textwidth}
    p{0.12\textwidth}p{0.12\textwidth}
    p{0.12\textwidth}p{0.12\textwidth} 
    p{0.1\textwidth}}
    \hline
    \multicolumn{7}{c}{Itex Domain B = 5\% of $\Omega$}\\
    \hline
      Agent &
      $R_{\it avg} (\downarrow)$  &  
      $P_{\it avg} (\downarrow)$  &  
      $U_{\it soc} (\uparrow)$  &  
      $U_{\it ind}^{total}  (\uparrow)$  &  
      $U_{\it ind}^s  (\uparrow)$  &
       $S_{\%} (\uparrow)$   \\  
      \hline
    
   ANESIA-DRL  & 555.81 $\pm$ 532.81 & \textbf{0.04 $\pm$ 0.16} & \textbf{1.27 $\pm$ 0.46} & 0.56 $\pm$ 0.12 & \textbf{0.91 $\pm$ 0.16} & 0.27 \\
   \hline\hline
    AgentGP $\bullet$ & 371.01 $\pm$ 177.31 & 0.12 $\pm$ 0.18 & 0.58 $\pm$ 0.38 & 0.51 $\pm$ 0.23 & 0.52 $\pm$ 0.27 & 0.77 \\
     FSEGA2019 $\bullet$ & 2714.98 $\pm$ 4328.78 & 0.11 $\pm$ 0.16 & 0.67 $\pm$ 0.37 & 0.66 $\pm$ 0.21 & 0.69 $\pm$ 0.21 & 0.82 \\
    \hline\hline
   AgentHerb $\diamond$  & \textbf{9.12 $\pm$ 6.25} & 0.05 $\pm$ 0.06 & 1.15 $\pm$ 0.08 & 0.26 $\pm$ 0.11 & 0.26 $\pm$ 0.11 & \textbf{1.00 }\\
    Agent33 $\diamond$ & 1850.84 $\pm$ 3677.84 & 0.07 $\pm$ 0.12 & 0.85 $\pm$ 0.34 & 0.42 $\pm$ 0.18 & 0.41 $\pm$ 0.18 & 0.92 \\
    Sontag $\diamond$  & 2965.81 $\pm$ 5008.5 & 0.14 $\pm$ 0.19 & 0.63 $\pm$ 0.43 & \textbf{0.69 $\pm$ 0.19} & 0.76 $\pm$ 0.17 & 0.73 \\
\hline
    \end{tabular}
    }
    \caption{Performance of ANESIA-DRL - Ablation Study1 - over domain Itex (1440 $\times 2$ profiles = 2880 simulations)}
    \label{table:ITEX}
\end{table*}

\begin{table*}
    \centering
    \resizebox{.98\textwidth}{!}{
    \begin{tabular}{p{0.18\textwidth} p{0.20\textwidth}
    p{0.12\textwidth}p{0.12\textwidth}
    p{0.12\textwidth}p{0.12\textwidth} 
    p{0.1\textwidth}}
    \hline
    \multicolumn{7}{c}{Outfit Domain B = 5\% of $\Omega$}\\
    \hline
      Agent &
      $R_{\it avg} (\downarrow)$  &  
      $P_{\it avg} (\downarrow)$  &  
      $U_{\it soc} (\uparrow)$  &  
      $U_{\it ind}^{total}  (\uparrow)$  &  
      $U_{\it ind}^s  (\uparrow)$  &
       $S_{\%} (\uparrow)$   \\  
      \hline
    ANESIA-DRL & 355.37 $\pm$ 256.24 & \textbf{0.05 $\pm$ 0.27} & \textbf{1.57 $\pm$ 0.67} & 0.76 $\pm$ 0.25 & \textbf{0.98 $\pm$ 0.06} & 0.54 \\
    \hline\hline
    AgentGP $\bullet$  & 704.03 $\pm$ 524.83 & 0.27 $\pm$ 0.27 & 0.71 $\pm$ 0.72 & 0.71 $\pm$ 0.22 & 0.91 $\pm$ 0.09 & 0.51 \\
    FSEGA2019 $\bullet$ & 690.15 $\pm$ 766.83 & 0.04 $\pm$ 0.14 & 1.27 $\pm$ 0.42 & 0.8 $\pm$ 0.15 & 0.83 $\pm$ 0.12 & 0.93 \\
    \hline\hline
    AgentHerb $\diamond$ & \textbf{4.04 $\pm$ 1.77} & \textbf{0.03 $\pm$ 0.07} & 1.55 $\pm$ 0.17 & 0.59 $\pm$ 0.18 & 0.59 $\pm$ 0.18 & \textbf{1.00 }\\
    Agent33 $\diamond$  & 244.91 $\pm$ 419.1 & 0.03 $\pm$ 0.1 & 1.47 $\pm$ 0.34 & 0.67 $\pm$ 0.14 & 0.67 $\pm$ 0.14 & 0.96 \\
    Sontag $\diamond$  & 600.69 $\pm$ 765.69 & 0.04 $\pm$ 0.14 & 1.32 $\pm$ 0.42 & 0.79 $\pm$ 0.14 & 0.81 $\pm$ 0.11 & 0.93 \\
\hline
    \end{tabular}
    }
    \caption{Performance of ANESIA-DRL - Ablation Study1 - over domain Outfit (1440 $\times 2$ profiles = 2880 simulations)}
    \label{table:OUTFIT}
\end{table*}

\begin{table*}
    \centering
    \resizebox{.98\textwidth}{!}{
    \begin{tabular}{p{0.18\textwidth} p{0.20\textwidth}
    p{0.12\textwidth}p{0.12\textwidth}
    p{0.12\textwidth}p{0.12\textwidth} 
    p{0.1\textwidth}}
    \hline
    \multicolumn{7}{c}{Airport Domain B = 5\% of $\Omega$}\\
    \hline
      Agent &
      $R_{\it avg} (\downarrow)$  &  
      $P_{\it avg} (\downarrow)$  &  
      $U_{\it soc} (\uparrow)$  &  
      $U_{\it ind}^{total}  (\uparrow)$  &  
      $U_{\it ind}^s  (\uparrow)$  &
       $S_{\%} (\uparrow)$   \\  
      \hline
    
    ANESIA-rand & 119.3 $\pm$ 125.27 & 0.2 $\pm$ 0.26 & 1.38 $\pm$ 0.38 & 0.6 $\pm$ 0.2 & 0.81 $\pm$ 0.2 & 0.95 \\
    \hline\hline
    AgentGP $\bullet$ & 127.66 $\pm$ 116.52 & 0.04 $\pm$ 0.14 & 1.48 $\pm$ 0.21 & 0.64 $\pm$ 0.12 & 0.65 $\pm$ 0.12 & 0.99 \\

    FSEGA2019 $\bullet$ & 439.89 $\pm$ 636.24 &\textbf{ 0.06 $\pm$ 0.14} & 1.47 $\pm$ 0.18 & 0.75 $\pm$ 0.16 & 0.75 $\pm$ 0.16 & \textbf{1.00} \\
      \hline\hline
        AgentHerb $\diamond$ & \textbf{10.64 $\pm$ 7.45} & 0.2 $\pm$ 0.22 & 1.25 $\pm$ 0.24 & 0.48 $\pm$ 0.22 & 0.48 $\pm$ 0.22 & \textbf{1.00} \\
    Agent33 $\diamond$  & 374.62 $\pm$ 500.5 & 0.2 $\pm$ 0.24 & 1.23 $\pm$ 0.35 & 0.54 $\pm$ 0.16 & 0.54 $\pm$ 0.16 & 0.99 \\
    Sontag $\diamond$ & 434.54 $\pm$ 488.25 & 0.11 $\pm$ 0.24 & 1.38 $\pm$ 0.34 & 0.73 $\pm$ 0.19 & 0.74 $\pm$ 0.19 & 0.97 \\
\hline
    \end{tabular}
    }
    \caption{Performance of \textit{ANESIA-Random} - Ablation Study2 - over domain AIRPORT (1440 $\times 2$ profiles = 2880 simulations)}
    \label{table:AIRPORT}
\end{table*}

\begin{table*}
    \centering
    \resizebox{.98\textwidth}{!}{
    \begin{tabular}{p{0.18\textwidth} p{0.20\textwidth}
    p{0.12\textwidth}p{0.12\textwidth}
    p{0.12\textwidth}p{0.12\textwidth} 
    p{0.1\textwidth}}
    \hline
    \multicolumn{7}{c}{Camera Domain B = 5\% of $\Omega$}\\
    \hline
      Agent &
      $R_{\it avg} (\downarrow)$  &  
      $P_{\it avg} (\downarrow)$  &  
      $U_{\it soc} (\uparrow)$  &  
      $U_{\it ind}^{total}  (\uparrow)$  &  
      $U_{\it ind}^s  (\uparrow)$  &
       $S_{\%} (\uparrow)$   \\  
      \hline
    
    ANESIA-rand & 274.92 $\pm$ 264.37 & 0.29 $\pm$ 0.09 &\textbf{ 1.19 $\pm$ 0.56} & 0.65 $\pm$ 0.15 & 0.71 $\pm$ 0.14 & 0.74 \\
    \hline\hline
    AgentGP $\bullet$ & 110.31 $\pm$ 240.73 & 0.16 $\pm$ 0.2 & 1.14 $\pm$ 0.42 & 0.63 $\pm$ 0.19 & 0.65 $\pm$ 0.2 & 0.91 \\

    FSEGA2019 $\bullet$  & 58.45 $\pm$ 37.13 & 0.16 $\pm$ 0.25 & 1.1 $\pm$ 0.49 & 0.78 $\pm$ 0.15 & 0.82 $\pm$ 0.1 & 0.86 \\
        \hline\hline

        AgentHerb $\diamond$ & \textbf{5.15 $\pm$ 2.07} & \textbf{0.06 $\pm$ 0.09} & 1.4 $\pm$ 0.21 & 0.52 $\pm$ 0.22 & 0.52 $\pm$ 0.22 & \textbf{1.00} \\
    Agent33 $\diamond$ & 410.81 $\pm$ 630.4 &\textbf{ 0.06 $\pm$ 0.07} & 1.4 $\pm$ 0.2 & 0.7 $\pm$ 0.18 & 0.7 $\pm$ 0.18 & \textbf{1.00} \\
    Sontag $\diamond$ & 651.37 $\pm$ 1116.79 & \textbf{0.06 $\pm$ 0.1 }& 1.36 $\pm$ 0.23 & 0.74 $\pm$ 0.14 & 0.74 $\pm$ 0.14 & 0.99 \\
\hline
    \end{tabular}
    }
    \caption{Performance of \textit{ANESIA-Random} - Ablation Study2 - over domain Camera (1440 $\times 2$ profiles = 2880 simulations) }
    \label{table:CAMERA}
\end{table*}

\begin{table*}
    \centering
    \resizebox{.98\textwidth}{!}{
    \begin{tabular}{p{0.18\textwidth} p{0.20\textwidth}
    p{0.12\textwidth}p{0.12\textwidth}
    p{0.12\textwidth}p{0.12\textwidth} 
    p{0.1\textwidth}}
    \hline
    \multicolumn{7}{c}{Energy Domain B = 5\% of $\Omega$}\\
    \hline
      Agent &
      $R_{\it avg} (\downarrow)$  &  
      $P_{\it avg} (\downarrow)$  &  
      $U_{\it soc} (\uparrow)$  &  
      $U_{\it ind}^{total}  (\uparrow)$  &  
      $U_{\it ind}^s  (\uparrow)$  &
       $S_{\%} (\uparrow)$   \\  
      \hline
    ANESIA-rand  & 824.42 $\pm$ 789.03 & 0.19 $\pm$ 0.21 & \textbf{1.11 $\pm$ 0.29} & 0.45 $\pm$ 0.22 & 0.46 $\pm$ 0.22 & 0.94 \\
    \hline\hline
    AgentGP $\bullet$ & 56.54 $\pm$ 208.7 & 0.47 $\pm$ 0.38 & 0.64 $\pm$ 0.55 & 0.42 $\pm$ 0.19 & 0.54 $\pm$ 0.15 & 0.58 \\
     FSEGA2019 $\bullet$ & 215.66 $\pm$ 125.98 & 0.6 $\pm$ 0.4 & 0.41 $\pm$ 0.53 & 0.48 $\pm$ 0.29 &\textbf{ 0.85 $\pm$ 0.08} & 0.38 \\
     \hline\hline
    AgentHerb $\diamond$ &\textbf{ 35.12 $\pm$ 25.49} & \textbf{0.08 $\pm$ 0.08} & 1.08 $\pm$ 0.1 & 0.25 $\pm$ 0.16 & 0.25 $\pm$ 0.16 & 1.00 \\
   
    Agent33 $\diamond$ & 7123.76 $\pm$ 9819.55 & 0.15 $\pm$ 0.12 & 1.10 $\pm$ 0.18 & 0.48 $\pm$ 0.17 & 0.48 $\pm$ 0.17 &\textbf{ 0.98 }\\
    Sontag $\diamond$  & 9379.17 $\pm$ 12799.92 & 0.37 $\pm$ 0.36 & 0.77 $\pm$ 0.51 & \textbf{0.57 $\pm$ 0.23} & 0.71 $\pm$ 0.12 & 0.71 \\
\hline
    \end{tabular}
    }
    \caption{Performance of \textit{ANESIA-Random} - Ablation Study2 - over domain Energy (1440 $\times 2$ profiles = 2880 simulations)}
    \label{table:ENERGY}
\end{table*}

\begin{table*}
    \centering
    \resizebox{.98\textwidth}{!}{
    \begin{tabular}{p{0.18\textwidth} p{0.20\textwidth}
    p{0.12\textwidth}p{0.12\textwidth}
    p{0.12\textwidth}p{0.12\textwidth} 
    p{0.1\textwidth}}
    \hline
    \multicolumn{7}{c}{Grocery Domain B = 5\% of $\Omega$}\\
    \hline
      Agent &
      $R_{\it avg} (\downarrow)$  &  
      $P_{\it avg} (\downarrow)$  &  
      $U_{\it soc} (\uparrow)$  &  
      $U_{\it ind}^{total}  (\uparrow)$  &  
      $U_{\it ind}^s  (\uparrow)$  &
       $S_{\%} (\uparrow)$   \\  
      \hline
    
    ANESIA-rand  & 237.6 $\pm$ 225.72 & \textbf{0.21 $\pm$ 0.24} & \textbf{1.27 $\pm$ 0.45} & 0.64 $\pm$ 0.28 & 0.73 $\pm$ 0.14 & 0.87 \\
        \hline\hline

    AgentGP $\bullet$ & 91.72 $\pm$ 175.14 & 0.21 $\pm$ 0.15 & \textbf{1.27 $\pm$ 0.31} & \textbf{0.72 $\pm$ 0.19} & \textbf{0.75 $\pm$ 0.13} & 0.96 \\
    FSEGA2019 $\bullet$ & 206.3 $\pm$ 292.85 & 0.22 $\pm$ 0.1 & 1.22 $\pm$ 0.22 & 0.64 $\pm$ 0.16 & 0.65 $\pm$ 0.14 & 0.98 \\
    \hline\hline
    AgentHerb $\diamond$ & \textbf{4.96 $\pm$ 2.37} & 0.26 $\pm$ 0.07 & 1.24 $\pm$ 0.14 & 0.51 $\pm$ 0.15 & 0.51 $\pm$ 0.15 & \textbf{1.00} \\
    Agent33 $\diamond$ & 90.53 $\pm$ 97.13 & 0.24 $\pm$ 0.07 & 1.25 $\pm$ 0.14 & 0.57 $\pm$ 0.16 & 0.57 $\pm$ 0.16 & \textbf{1.00} \\
    Sontag $\diamond$  & 363.54 $\pm$ 669.3 & 0.21 $\pm$ 0.08 & 1.25 $\pm$ 0.17 & 0.66 $\pm$ 0.14 & 0.66 $\pm$ 0.13 & 0.99 \\
\hline
    \end{tabular}
    }
    \caption{Performance of \textit{ANESIA-Random} - Ablation Study2 - over domain Grocery (1440 $\times 2$ profiles = 2880 simulations)}
    \label{table:AIRPORT}
\end{table*}

\begin{table*}
    \centering
    \resizebox{.98\textwidth}{!}{
 \begin{tabular}{p{0.18\textwidth} p{0.20\textwidth}
    p{0.12\textwidth}p{0.12\textwidth}
    p{0.12\textwidth}p{0.12\textwidth} 
    p{0.1\textwidth}}
    \hline
    \multicolumn{7}{c}{Fitness Domain B = 5\% of $\Omega$}\\
    \hline
      Agent &
      $R_{\it avg} (\downarrow)$  &  
      $P_{\it avg} (\downarrow)$  &  
      $U_{\it soc} (\uparrow)$  &  
      $U_{\it ind}^{total}  (\uparrow)$  &  
      $U_{\it ind}^s  (\uparrow)$  &
       $S_{\%} (\uparrow)$   \\  
      \hline
    
    ANESIA-rand & 7.11 $\pm$ 2.59 & 0.05 $\pm$ 0.07 &\textbf{ 1.49 $\pm$ 0.08} & 0.89 $\pm$ 0.13 & \textbf{0.89 $\pm$ 0.13} & \textbf{1.00} \\
        \hline\hline

    AgentGP $\bullet$ & 577.86 $\pm$ 1122.43 & 0.19 $\pm$ 0.31 & 1.16 $\pm$ 0.56 & 0.64 $\pm$ 0.16 & 0.67 $\pm$ 0.16 & 0.82 \\
    FSEGA2019 $\bullet$  & 1476.17 $\pm$ 1984.41 & 0.1 $\pm$ 0.16 & 1.3 $\pm$ 0.29 & 0.77 $\pm$ 0.13 & 0.78 $\pm$ 0.12 & 0.96 \\
        \hline\hline

    AgentHerb $\diamond$  & \textbf{7.23 $\pm$ 3.22} & \textbf{0.04 $\pm$ 0.05} & \textbf{1.49 $\pm$ 0.05} & 0.58 $\pm$ 0.1 & 0.58 $\pm$ 0.1 & \textbf{1.00} \\
    
    Agent33 $\diamond$ & 569.85 $\pm$ 956.24 & 0.07 $\pm$ 0.05 & 1.43 $\pm$ 0.07 & 0.69 $\pm$ 0.12 & 0.69 $\pm$ 0.12 & \textbf{1.00} \\
    Sontag $\diamond$  & 1708.57 $\pm$ 2728.7 & 0.09 $\pm$ 0.14 & 1.33 $\pm$ 0.26 & 0.78 $\pm$ 0.14 & 0.78 $\pm$ 0.13 & 0.97 \\
\hline
    \end{tabular}
    }
    \caption{Performance of \textit{ANESIA-Random} - Ablation Study2 - over domain Fitness (1440 $\times 2$ profiles = 2880 simulations)}
    \label{table:AIRPORT}
\end{table*}

\begin{table*}
    \centering
    \resizebox{.98\textwidth}{!}{
    \begin{tabular}{p{0.18\textwidth} p{0.20\textwidth}
    p{0.12\textwidth}p{0.12\textwidth}
    p{0.12\textwidth}p{0.12\textwidth} 
    p{0.1\textwidth}}
    \hline
    \multicolumn{7}{c}{Flight Booking Domain B = 5\% of $\Omega$}\\
    \hline
      Agent &
      $R_{\it avg} (\downarrow)$  &  
      $P_{\it avg} (\downarrow)$  &  
      $U_{\it soc} (\uparrow)$  &  
      $U_{\it ind}^{total}  (\uparrow)$  &  
      $U_{\it ind}^s  (\uparrow)$  &
       $S_{\%} (\uparrow)$   \\  
      \hline
    
    ANESIA-rand & 579.56 $\pm$ 684.55 & 0.34 $\pm$ 0.27 & \textbf{1.16 $\pm$ 0.63} & \textbf{0.66 $\pm$ 0.21} & 0.77 $\pm$ 0.24 & 0.46 \\
    \hline\hline
    AgentGP $\bullet$ & 529.11 $\pm$ 275.89 & 0.43 $\pm$ 0.24 & 0.38 $\pm$ 0.57 & 0.55 $\pm$ 0.12 & 0.65 $\pm$ 0.16 & 0.32 \\
    FSEGA2019 $\bullet$ & 3146.95 $\pm$ 4221.66 & 0.46 $\pm$ 0.24 & 0.24 $\pm$ 0.49 & 0.60 $\pm$ 0.2 & \textbf{1.0 $\pm$ 0.0} & 0.21 \\
        \hline\hline

    AgentHerb $\diamond$ & \textbf{553.35 $\pm$ 1149.49} & \textbf{0.12 $\pm$ 0.2} & \textbf{1.16 $\pm$ 0.47} & 0.52 $\pm$ 0.16 & 0.52 $\pm$ 0.17 & 0.87 \\
    Agent33 $\diamond$ & 1633.69 $\pm$ 2938.1 & \textbf{0.12 $\pm$ 0.2} & 1.01 $\pm$ 0.43 & 0.46 $\pm$ 0.15 & 0.45 $\pm$ 0.16 & \textbf{0.87} \\
    Sontag $\diamond$  & 3183.22 $\pm$ 4665.68 & 0.38 $\pm$ 0.27 & 0.48 $\pm$ 0.62 & \textbf{0.66 $\pm$ 0.21} & 0.91 $\pm$ 0.11 & 0.39 \\
\hline
    \end{tabular}
    }
    \caption{Performance of \textit{ANESIA-Random} - Ablation Study2 - over domain Flight Booking (1440 $\times 2$ profiles = 2880 simulations)}
    \label{table:AIRPORT}
\end{table*}

\begin{table*}
    \centering
    \resizebox{.98\textwidth}{!}{
   \begin{tabular}{p{0.18\textwidth} p{0.20\textwidth}
    p{0.12\textwidth}p{0.12\textwidth}
    p{0.12\textwidth}p{0.12\textwidth} 
    p{0.1\textwidth}}
    \hline
    \multicolumn{7}{c}{Itex Domain B = 5\% of $\Omega$}\\
    \hline
      Agent &
      $R_{\it avg} (\downarrow)$  &  
      $P_{\it avg} (\downarrow)$  &  
      $U_{\it soc} (\uparrow)$  &  
      $U_{\it ind}^{total}  (\uparrow)$  &  
      $U_{\it ind}^s  (\uparrow)$  &
       $S_{\%} (\uparrow)$   \\  
      \hline
    
    ANESIA-rand & 617.8 $\pm$ 561.02 & 0.14 $\pm$ 0.2 & 1.13 $\pm$ 0.44 & 0.67 $\pm$ 0.18 & 0.74 $\pm$ 0.22 & 0.53 \\
    \hline\hline
    AgentGP $\bullet$ & 378.94 $\pm$ 185.01 & 0.11 $\pm$ 0.17 & 0.6 $\pm$ 0.36 & 0.53 $\pm$ 0.24 & 0.53 $\pm$ 0.26 & 0.80 \\
      FSEGA2019 $\bullet$ &  2987.34 $\pm$ 3904.42 & 0.1 $\pm$ 0.15 & 0.68 $\pm$ 0.35 & 0.66 $\pm$ 0.2 & 0.68 $\pm$ 0.21 & 0.85 \\
    \hline\hline
    AgentHerb $\diamond$ &  \textbf{9.19 $\pm$ 5.43} & \textbf{0.04 $\pm$ 0.06} & \textbf{1.14 $\pm$ 0.09} & 0.26 $\pm$ 0.11 & 0.26 $\pm$ 0.11 & \textbf{1.00} \\
  
    Agent33 $\diamond$ & 1731.26 $\pm$ 2910.4 & 0.06 $\pm$ 0.1 & 0.87 $\pm$ 0.3 & 0.41 $\pm$ 0.18 & 0.41 $\pm$ 0.18 & 0.96 \\
    Sontag $\diamond$ &  2993.49 $\pm$ 4288.98 & 0.13 $\pm$ 0.18 & 0.65 $\pm$ 0.41 & \textbf{0.69 $\pm$ 0.19 }& 0.75 $\pm$ 0.18 & 0.77 \\
\hline
    \end{tabular}
    }
    \caption{Performance of \textit{ANESIA-Random} - Ablation Study2 - over domain Itex (1440 $\times 2$ profiles = 2880 simulations)}
    \label{table:AIRPORT}
\end{table*}

\begin{table*}
    \centering
    \resizebox{.98\textwidth}{!}{
    \begin{tabular}{p{0.18\textwidth} p{0.20\textwidth}
    p{0.12\textwidth}p{0.12\textwidth}
    p{0.12\textwidth}p{0.12\textwidth} 
    p{0.1\textwidth}}
    \hline
    \multicolumn{7}{c}{Outfit Domain B = 5\% of $\Omega$}\\
    \hline
      Agent &
      $R_{\it avg} (\downarrow)$  &  
      $P_{\it avg} (\downarrow)$  &  
      $U_{\it soc} (\uparrow)$  &  
      $U_{\it ind}^{total}  (\uparrow)$  &  
      $U_{\it ind}^s  (\uparrow)$  &
       $S_{\%} (\uparrow)$   \\  
      \hline
    
    ANESIA-rand & 231.72 $\pm$ 201.41 & 0.16 $\pm$ 0.25 & 1.16 $\pm$ 0.61 & \textbf{0.91 $\pm$ 0.22} & \textbf{0.95 $\pm$ 0.1} & 0.70 \\
            \hline\hline

    AgentGP $\bullet$ & 434.14 $\pm$ 336.49 & 0.17 $\pm$ 0.25 & 0.85 $\pm$ 0.64 & 0.78 $\pm$ 0.2 & 0.91 $\pm$ 0.08 & 0.69 \\
    FSEGA2019 $\bullet$ & 329.94 $\pm$ 300.72 & \textbf{0.02 $\pm$ 0.1 }& 1.29 $\pm$ 0.35 & 0.81 $\pm$ 0.14 & 0.82 $\pm$ 0.12 & 0.96 \\
    \hline\hline

    AgentHerb $\diamond$ & \textbf{4.03 $\pm$ 1.69} & 0.03 $\pm$ 0.08 &\textbf{ 1.54 $\pm$ 0.17} & 0.59 $\pm$ 0.19 & 0.59 $\pm$ 0.19 & \textbf{1.00} \\
   
    Agent33 $\diamond$ & 133.28 $\pm$ 184.98 & 0.03 $\pm$ 0.1 & 1.45 $\pm$ 0.34 & 0.66 $\pm$ 0.15 & 0.67 $\pm$ 0.15 & 0.97 \\
    Sontag $\diamond$  & 330.15 $\pm$ 416.61 & \textbf{0.02 $\pm$ 0.11} & 1.33 $\pm$ 0.37 & 0.79 $\pm$ 0.13 & 0.81 $\pm$ 0.11 & 0.96 \\
\hline
    \end{tabular}
    }
    \caption{Performance of \textit{ANESIA-Random} - Ablation Study2 - over domain Outfit (1440 $\times 2$ profiles = 2880 simulations)}
    \label{table:Outfit}
\end{table*}

\end{document}